%% file: MAIN.tex
\documentclass[authoryear,12pt]{elsarticle}

\usepackage{amssymb}
\usepackage{amsmath}

\usepackage{graphicx} 
\usepackage{amsmath}
\usepackage{amsthm}
\newtheorem{lemma}{Lemma}
\usepackage{amsthm}
\usepackage{graphicx}
\usepackage{subcaption}
\usepackage{color}
\usepackage{mwe}
\usepackage{hyperref}
\usepackage{url}

\usepackage{ltablex,array}
\usepackage{longtable}
\usepackage{amssymb}
\usepackage{siunitx}
\usepackage{amsmath}
\usepackage{arydshln}
\usepackage{multirow}
\usepackage{booktabs}
\usepackage{hyperref}
\usepackage{graphicx} 
\usepackage{nameref}
\usepackage{wrapfig}
\usepackage{tabularx}
\usepackage[ruled,vlined,linesnumbered]{algorithm2e}
\usepackage{xcolor}
\usepackage{tcolorbox} 
\tcbuselibrary{skins} 
\usepackage{mathtools}
\usepackage{adjustbox}
\usepackage{graphicx}
\usepackage{booktabs}
\usepackage[capitalize]{cleveref}
\usepackage{dot2texi}
\usepackage{tikz}
\usepackage{amsmath}
\usetikzlibrary{shapes,arrows,arrows.meta,positioning} 
\usetikzlibrary{arrows.meta,graphs}

\newcommand\blockSet{{\mathcal{B}}}
\newcommand\block[1]{{{B}^{#1}}}
\newcommand\blockInstanceSet[1]{{\mathcal{I}^{#1}}}
\newcommand\blockInstance[2]{{I^{#1}_{#2}}}
\newcommand\rectangleSet[1]{\mathcal{R}^{#1}}
\newcommand\rectangle[2]{{R^{#1}_{#2}}}
\newcommand\blockToInstance[2]{\sigma(\blockInstance{#1}{#2})}
\newcommand\nmRectangle[1]{{m_{#1}}}
\newcommand\nmInstance[1]{{n_{#1}}}
\newcommand\width[1]{W^{#1}}
\newcommand\height[1]{H^{#1}}
\newcommand\object[2]{{O^{#1}_{#2}}}
\newcommand\objectSet[1]{{\mathcal{O}^{#1}}}
\newcommand\LBarea[1]{LB_{\texttt{area}}(#1)}
\newcommand\LBhp[1]{LB_{\texttt{W+H}}(#1)}
\newcommand\LBareatop[0]{LB_{\texttt{area}}}
\newcommand\LBhptop[0]{LB_{\texttt{W+H}}}

\journal{Computers \& Operations Research}

\begin{document}

\begin{frontmatter}




\title{Hierarchical Rectangle Packing Solved by Multi-Level Recursive Logic-based Benders Decomposition}



\cortext[corr1]{Corresponding author}

\affiliation[dcefel]{organization={DCE, FEE, Czech Technical University in Prague}, city={Praha}, country={Czech Republic}}
\affiliation[ciirc]{organization={IID, CIIRC, Czech Technical University in Prague}, city={Praha}, country={Czech Republic},}
\affiliation[laas]{organization={LAAS-CNRS, Université de Toulouse, CNRS}, city={Toulouse}, country={France},}

\author[dcefel,ciirc]{Josef Grus\corref{corr1}}
\ead{josef.grus@cvut.cz}
\author[ciirc]{Zdeněk Hanzálek}
\ead{zdenek.hanzalek@cvut.cz}
\author[laas]{Christian Artigues}
\ead{christian.artigues@laas.fr}
\author[laas]{Cyrille Briand}
\ead{cyrille.briand@laas.fr}
\author[laas]{Emmanuel Hebrard}
\ead{emmanuel.hebrard@laas.fr}

\begin{abstract}
We study the two-dimensional hierarchical rectangle packing problem, motivated by applications in analog integrated circuit layout, facility layout, and logistics. Unlike classical strip or bin packing, the dimensions of the container are not fixed, and the packing is inherently hierarchical: each item is either a rectangle or a block occurrence, whose dimensions are a solution of another packing problem. This recursive structure reflects real-world scenarios in which components, boxes, or modules must be packed within higher-level containers. We formally define the problem and propose exact formulations in Mixed-Integer Linear Programming and Constraint Programming. Given the computational difficulty of solving complex packing instances directly, we propose decomposition heuristics. First, we implement an existing Bottom-Up baseline method that solves subblocks before combining them at higher levels. Building upon this, we introduce a novel multilevel Logic-based Benders Decomposition method. This heuristic method dynamically refines dimension constraints of block types, eliminating the need for manual selection of candidate widths or aspect ratios. Experiments on synthetic instances with up to seven hierarchy levels, 80 items per block type, and limited computation time show that the proposed decomposition significantly outperforms both monolithic formulations and the Bottom-Up method in terms of solution quality and scalability.
\end{abstract}


\begin{keyword}
recursive decomposition \sep
rectangle packing \sep
constraint programming \sep
hierarchical packing
\end{keyword}
\end{frontmatter}

\input{sections/1-introduction}
\input{sections/2-related-work}
\input{sections/3-problem-statement}
\input{sections/4-decomposition}

\input{sections/4b-decomposition}
\input{sections/5-experiments}
\input{sections/6-discussion-conclusion}

\section*{Acknowledgments}
This work was co-funded by the European Union under the project ROBOPROX (reg. no. CZ.02.01.01/00/22\_008/0004590) and by the  Artificial and Natural Intelligence Toulouse Institute (ANITI) under the grant agreement ANR-23-IACL-0002. 

\bibliographystyle{elsarticle-num-names} 
\bibliography{biblio}
\end{document}

%% file: sections/1-introduction.tex
\section{Introduction}\label{sec:introduction}
In this paper, we focus on the two-dimensional rectangle packing problem inspired by the placement of components of analog integrated circuits. Our objective is to find the smallest rectangular container (called a block type, as it serves as a sort of template) that can contain the rectangular items so that they do not overlap. The width and height of the block type are to be decided, unlike in the case of strip packing (where the container has a fixed width) or bin packing (multiple bins with fixed dimensions). In this paper, the packing problem itself is hierarchical. This means that an item to be packed is either an individual rectangle with fixed dimensions or an occurrence of another block type, meaning that the optimized block type contains a solution to another packing problem, whose solution serves as a template. To find the dimensions of such a block occurrence, it is necessary to recursively pack such a subblock type, which itself consists of individual rectangles and possibly other block occurrences. This hierarchy of block types is a static property of the problem instance, and can be conveniently represented as an out-tree, as \cref{fig:example-hierarchy} later demonstrates.

In packing applications, such a hierarchy is often an intrinsic property of the problem. In the design of integrated circuits, engineers also solve such a hierarchical packing problem. They need to design lower-level components, such as operational amplifiers, so that when they are used as parts of more complex components, they can be efficiently put together and the overall area of the circuit is minimized (ensuring more circuits can fit the wafer they are fabricated on). Individual components need to be encapsulated and isolated, with well-defined boundaries. Therefore, the problem cannot be solved as a single-block packing problem, and the hierarchy needs to be explicitly considered.

Another example of the application of the hierarchical packing problem is logistics. Assume that goods have the same height, so the problem essentially boils down to the two-dimensional case. Individual items are first grouped into boxes (e.g., per customer), which can then be packed into larger boxes, and finally loaded into a main container. This naturally leads to a multi-level hierarchical packing structure. However, in many logistics settings, at least one dimension of the main container is fixed. In such cases, the problem is more appropriately modeled using a strip-packing objective at the top level, rather than unconstrained area minimization.

In this paper, we formally describe the problem of hierarchical packing of two-dimensional items with an arbitrary number of levels, which, to the best of our knowledge, has not been formally studied in the literature. In \cref{sec:related-work}, related work on topics of packing, its applications, and the use of decomposition methods is investigated. In \cref{sec:desc}, the problem is formally described, and models for Mixed-Integer Linear Programming (MILP) and Constraint Programming (CP) solvers are proposed. In \cref{sec:baselines}, heuristics for finding initial solutions and the baseline Bottom-Up method are outlined. In \cref{sec:benders}, we develop a heuristic method based on Logic-based Benders Decomposition (LBBD) to alleviate the natural hierarchical property of the problem. Experiments in \cref{sec:experiments} show that the proposed method outperforms both exact monolithic models and the baseline Bottom-Up method in synthetic instances with up to 7 levels of hierarchy, with up to 80 items per block type, and limited computation time. Finally, in \cref{sec:discussion,sec:conclusion}, we elaborate on properties of the hierarchical packing problem and possible generalizations to industrial applications.

%% file: sections/2-related-work.tex
\section{Related Work}\label{sec:related-work}

Three research domains are directly relevant to this work: (i)~2D rectangle packing and cutting, (ii)~hierarchical and multi-level optimization, and (iii)~decomposition approaches, especially nested ones. While each domain is mature in isolation, their intersection (hierarchical 2D rectangle packing solved via nested/recursive decomposition) has not been systematically studied. This section investigates each research domain and relates this paper to it.

\subsection{Packing and Cutting}

Packing and cutting problems belong to the fundamental topics in operations research, which brought up many decomposition approaches, such as column generation \citep{gomory-1d,gomory-2d}. These problems have been extensively studied in the past. The recent survey paper by \citet{altogether-survey} serves as an overall introduction to the 2D case. Surveys by \citet{exact-survey,heur-survey} focus on exact approaches and heuristics, respectively.
 
MILP is often an approach of choice to tackle rectangle packing problems. Space-indexed (or grid-indexed) models, similar to the time-indexed formulations found in the scheduling domain, were utilized for 2D \citep{space-index-beasley} and 3D \citep{space-indexed-marecek} packing problems. These models have a pseudo-polynomial number of variables but are known to offer a good linear relaxation. ``Normal patterns'', a reduced set of relevant packing solutions, were successfully used to reduce the number of variables. Polynomial-sized relative-position models were described in \citet{relative-container} and found application in many CP-based approaches \citep{pre-korf,Korf2010}, as well as in this paper.  Their use in an MILP context is problematic due to the necessity of using ``big-M'' coefficients to encode the relative position constraints between rectangles.
 
Classical Benders decomposition was used in \citet{ZHANG2025} to remove big-M coefficients from the relative-position model for the integrated circuit placement. The Benders decomposition with combinatorial cuts was used in \citet{cote-cuts} for strip packing and later in \citet{cote-cuts2} for 2D bin packing; in both papers, the master problem exploits a relaxation through contiguous 1D bin packing \citep{logic-based-graph}. Column generation and branch-and-price were used mostly for packing 2D objects into the smallest number of bins \citep{pisinger-binpacking,cintra-columns}, where column generation was additionally applied at multiple levels to solve the guillotine cutting in \citet{cintra-columns} and nested column generation was studied in \citet{nested-column-cutting}. As was done in 1D bin packing \citep{temp-1d-binpack,lewis-trapezoid}, the pricing problem was solved using CP and dynamic programming, respectively.
 
Efficient lower bounds are crucial. Early work for strip packing was done in \citet{martello-lb}; \citet{alvarez-lb} and \citet{Boschetti-lb} extended these results and developed new methods that improve on the trivial area bound.  More advanced methods mostly assume fixed orientations; the free-orientation case was investigated in \citet{logic-based-graph}.  Pre-processing techniques that reduce the strip width or increase item widths were developed in \citet{Boschetti-lb}. A method for finding the minimum enclosing square was developed in \citet{martello-minimum-square}. Despite this rich literature, all of the above methods address a flatpacking problem: a single level of items packed into a single container.

2D packing problems arise in many application domains that naturally exhibit a hierarchical structure, yet that structure has been handled only informally. In integrated-circuit design \citep{xu-placement,zhu-circuits,grus-placement}, the goal is the smallest placement of components subject to connectivity constraints. In \citet{xu-placement} the circuit hierarchy is considered and solved in a bottom-up manner, which serves as a baseline in this paper (\cref{sec:bottom-up}). The facility layout problem \citep{kubalik,facility-exact} packs departments or rooms into a building to minimize material flow. The two-level case in \citet{facility-exact} contains a special case of the problem studied here, solved via a monolithic mathematical programming model. The relationship between periodic scheduling and packing \citep{grus-perioic} connects the two classical research domains, and scheduling with uncertain processing times was transformed into packing of ``F-shapes'' in \citet{NOVAK2019687}. 

Altogether, 2D packing is a well-researched topic that (with the exception of \cite{facility-exact}) handles hierarchical packing via heuristic strategies, such as bottom-up, or not at all. This paper aims to investigate this extension of rectangle packing in a systematic way.

\subsection{Hierarchical Optimization}

Problems with a hierarchical or recursive structure have been investigated more formally in other domains. Hierarchical planning is an extension of classical planning with task hierarchy \citep{hierarchical-planning-survey}; task hierarchy can model problem physics and guide the search. The two-level vehicle routing problem was studied in \citet{heur-benders-1,heur-benders-2}, solved using decomposition and heuristics. Hierarchical scheduling is prevalent in edge computing, where resources \citep{fog-hierarchy} or schedulers \citep{edge-hierarchy} are organized into a capability hierarchy. Problems with hierarchical or recursive structure similar to ours have also been investigated in the context of bilevel optimization \citep{bilevelsurvey}.
 
In the bin packing domain specifically, \citet{data-drive-multi-elvel} and \citet{Blansch_2022_bachelor} tackled a 1D multi-level bin packing case. The authors packed items into low-level bins that must, in turn, fit into upper-level bins using methods designed to generalize to an arbitrary number of levels. These works are the most direct precursors to ours. However, the one-dimensional setting avoids the difficulty of 2D packing itself: deciding whether a set of rectangles fits a container is already NP-hard, and this subproblem must be solved efficiently at every level of the hierarchy.

\subsection{Nested Decomposition}
 
Decomposition methods are sometimes applied in a nested (or, in a special case, recursive) manner. Nested branch-and-price was used to solve vehicle routing with complex inter-dependencies in \citet{nested-column-vrp}.  Nested logic-based Benders decomposition (LBBD) was used to plan first global and then local disaster response in \citet{nested-bd-disaster}, and for home healthcare planning in \citet{nested-cyril}. Benders decomposition with two levels was combined with dynamic programming in \citet{mo-nested-bd}. These works demonstrate that nested decomposition is a viable strategy for multi-level problems, but none of them involve a 2D packing subproblem, which is the topic of this paper.
 
\subsection{Contributions}
In summary, three mature research domains were discussed:
(i)~powerful exact methods for flat 2D rectangle packing;
(ii)~multi-level bin packing restricted to one dimension; and
(iii)~nested decomposition methods for problems other than 2D packing.
This paper sits at the intersection of the three domains and makes the following advances beyond the existing literature:
\begin{itemize}
  \item We formally define 2DHRP as a generalization of work shown in \citep{facility-exact}, and establish its relationship to flat 2D packing problems.
  \item We propose a recursive decomposition method that propagates information both top-down and bottom-up across hierarchy levels, improving the performance in contrast to bottom-up methods
        \citep{xu-placement}.
  \item We provide a comprehensive computational study on newly introduced benchmark instances, demonstrating the practical effectiveness of the approach.
\end{itemize}

%% file: sections/3-problem-statement.tex
\section{Problem Description}\label{sec:desc}
\subsection{Problem Statement}
In this section, we formally describe the 2D hierarchical rectangle packing (2DHRP) problem derived from application-specific domains \citep{xu-placement,facility-exact}. As outlined in \cref{sec:introduction}, the 2DHRP problem packs rectangles into lower-level rectangular block types, which each need to be packed into upper-level rectangular block types together with the respective individual rectangles of that level. 

\begin{figure}
        \centering
        \begin{subfigure}[b]{0.7\textwidth}  
            \input{sections/graph.tex}
            \caption{Hierarchy with four block types and eight rectangles.} 
            \label{fig:example-hierarchy}
            
        \end{subfigure}

    \vskip\baselineskip
    
        \begin{subfigure}[b]{0.7\textwidth}
            \centering
            \includegraphics[width=\textwidth]{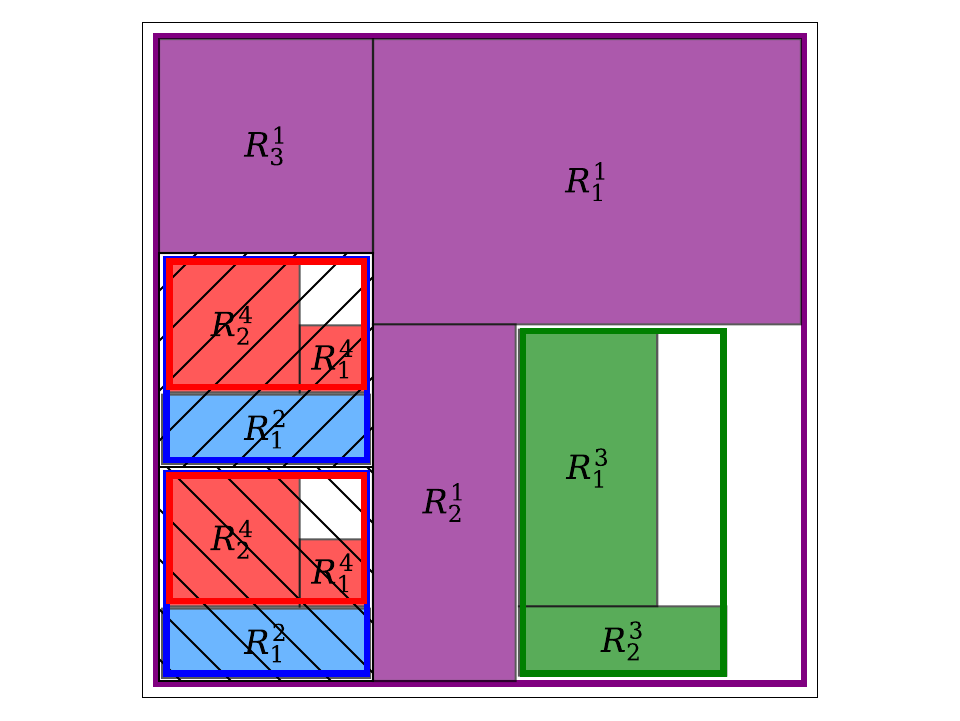}
            \caption{Optimal solution. Notice, that two occurrences $\blockInstance{1}{1},\blockInstance{1}{2}$ of block type $\block{2}$ (i.e., $\blockToInstance{1}{1}=\blockToInstance{1}{2}=2$), included in the top block type $\block{1}$, are shown in the bottom-left part of the figure, each with different hatching style.}    
            \label{fig:example-solution}
        \end{subfigure}
        \caption{Example instance of 2DHRP with an optimal solution. Four block types are organized in a weighted out-tree as shown in \cref{fig:example-hierarchy}. There are two block occurrences of block type $\block{2}$ as part of $\block{1}$.} 
        \label{fig:example-instance}
\end{figure}

The problem instance can be visualized as a weighted directed out-tree in \cref{fig:example-hierarchy}. Each block type $\block{i}$ is represented by a larger labeled node. The rectangles $\rectangle{i}{j}$ of the block type are represented by the smaller leaf nodes connected to the block type nodes (having the same color). Root node ($\block{1}$ in \cref{fig:example-hierarchy}) and its corresponding block type are referred to as a top node and top block type. An edge between two nodes indicates that the parent block type contains the corresponding child block type or rectangle. Since the hierarchy is assumed to be a tree, there exists exactly one path from the root to any node. Consequently, each block type, except for the root, is contained in exactly one parent block type. The number of occurrences of a child block type or rectangle is specified by the weight of the corresponding edge.

Crucially, all individual occurrences of the same block type are packed in the same way; this constraint originates from the circuit design domain, which requires the reuse of designed components, which are represented by block types. This also means that each occurrence of the block has the same dimensions. Unlike the block occurrences, each rectangle is independent of the others. Thus, to model two rectangles with the same dimensions, they would be represented by two distinct leaf nodes (i.e., their incoming edges have weight equal to 1).

Formally, let $\blockSet=\left\{\block{1},\dots,\block{i},\dots,\block{n}\right\}$ be a set of block types. Each block type $\block{i}$ consists of a set of its own independent rectangles $\rectangleSet{i}=\left\{\rectangle{i}{1},\dots,\rectangle{i}{j}\dots,\rectangle{i}{\nmRectangle{i}}\right\}$ and a set of block occurrences $\blockInstanceSet{i} = \left\{\blockInstance{i}{1},\dots,\blockInstance{i}{k},\dots,\blockInstance{i}{\nmInstance{i}}\right\}$. Each block occurrence $\blockInstance{i}{k}$ represents an instantiation of some block type $\block{i'}$ within the parent block type $\block{i}$. A given block type may be instantiated multiple times within another block type. The mapping:
\begin{equation}
    \sigma: \cup_i\blockInstanceSet{i} \rightarrow \left\{1,\dots,n\right\}
\end{equation} thus associates the block occurrence $\blockInstance{i}{k}$ to its block type template $\block{i'}$, $\blockToInstance{i}{k} = i'$. The values of $\blockToInstance{i}{k}$ for $\block{i}$ also determine which block types are directly included in $\block{i}$. Any block type with index from set $\mathcal{C}^i=\left\{i'~| ~\exists \blockInstance{i}{k}: \blockToInstance{i}{k} = i' \right\}$ is a child of $\block{i}$. This induces an edge from node $\block{i}$ to $\block{i'}$ in the hierarchy graph. This mapping between the hierarchy of the instance and the associated packing is highlighted in \cref{fig:example-solution}.


The main task is to minimize the size of the top block type while respecting packing constraints across the hierarchy. Each block type $\block{i}$ needs to be assigned its integer width and height $\width{i},\height{i}$, so no two objects $\object{i}{u},\object{i}{v}$ of the object set $\objectSet{i} = \blockInstanceSet{i} \cup \rectangleSet{i}$ overlap and are all are within the boundaries of the block type given by its dimensions $\width{i},\height{i}$. This is expressed by the constraints:
\begin{align}
    &(0 \le x_\object{i}{u})~\wedge~ (x_\object{i}{u} + w_\object{i}{u} \le W^i) \label{eq:a1}\\
    &(0 \le y_\object{i}{u})~\wedge~ (y_\object{i}{u} + h_\object{i}{u} \le H^i)\label{eq:a2}\\
    & (x_\object{i}{u} + w_\object{i}{u} \le  x_\object{i}{v}) \vee (x_\object{i}{v} + w_\object{i}{v} \le  x_\object{i}{u}) ~\vee \nonumber\\
    & (y_\object{i}{u} + h_\object{i}{u} \le  y_\object{i}{v}) \vee (y_\object{i}{v} + h_\object{i}{v} \le  y_\object{i}{u})\label{eq:a3}
\end{align}

where $(x,y)$ are the integer coordinates of the bottom-left corner of the rectangle or block occurrence, and $(w,h)$ are its integer width and height. 

When a rectangle $\rectangle{i}{j}$ is packed, its dimensions are chosen from a set $\mathcal{D}^i_j$ of available pre-defined variants:
\begin{equation}
    \mathcal{D}^i_j =\left\{(w^i_{j,1}, h^i_{j,1}),\dots,(w^i_{j,|\mathcal{D}^i_j|}, h^i_{j,|\mathcal{D}^i_j|})\right\}
\end{equation}
where we assume both the width and height of each variant are integers (ensuring that the coordinates of rectangles will be integers as well). Exactly one of the $|\mathcal{D}^i_j|$ available variants has to be selected per rectangle. Note that the rotation of a single-variant rectangle is modeled by adding another reflected variant to $\mathcal{D}^i_j$. 

On the other hand, the dimensions of the block occurrence $\blockInstance{i}{k}$ are given by the packing of its reference block type $\block{i'},i'=\blockToInstance{i}{k}$. This means that $w_\blockInstance{i}{k} = \width{i'},h_\blockInstance{i}{k} = \height{i'}$. A packing specifying coordinates and dimensions for each rectangle, block occurrence, and dimensions of each block type, which respects \cref{eq:a1,eq:a2,eq:a3}, is a feasible solution of the 2DHRP problem.  

The objective is to minimize the size of the top block type (w.l.o.g., the top block type can be assumed to be $\block{1}$). Let its dimensions be $\width{}=\width{1},\height{}=\height{1}$. Although minimization of area $\width{}\cdot\height{}$ is a natural objective, the proxy criterion of half-perimeter $\width{}+\height{}$ is minimized instead. From previous work \citep{xu-placement} and preliminary experiments, given a limited computation time, optimization of the half-perimeter proxy seems to be more efficient even with respect to the final area. Furthermore, the square-like solutions this objective prefers are of interest for both the logistics and circuit design applications. Naturally, when a case with one dimension fixed is encountered, the other dimension is directly minimized in a strip packing manner.

An illustration of the problem is shown in \cref{fig:example-instance}. The colors of the rectangles correspond to the color of their block type's node in the hierarchy. We can see that the block type $\block{2}$ was used twice (blue boundary) and consists of a single blue rectangle $\rectangle{2}{1}$ and a block occurrence of red block type $\block{4}$. For the top block type $\block{1}$, we need to pack the two (blue) block occurrences of $\block{2}$ with a single (green) occurrence of $\block{3}$ and three purple rectangles $\rectangle{1}{1},\rectangle{1}{2},\rectangle{1}{3}$.

\subsection{MILP Model}\label{sec:milp}

We first formulate the 2DHRP problem using an MILP model. A related multiple-block-type relative-position-based formulation was used by \citet{facility-exact}, where the authors simultaneously optimized the layout of the plant as well as a processing equipment room within the plant. We generalize the approach to handle an arbitrary number of block types. For block type $\block{i}, i \in\left\{1,\dots,n\right\}$, the partial model is the following:
\begingroup
\allowdisplaybreaks
\begin{align}
    &0 \le x_\object{i}{u} \le W^i - w_\object{i}{u} & \forall \object{i}{u} \in \objectSet{i} \label{eq:11}\\
    &0 \le y_\object{i}{u} \le H^i -  h_\object{i}{u} & \forall \object{i}{u} \in \objectSet{i} \label{eq:12}\\
    & \sum_{k=1}^{4} r^{i,k}_{u,v} \ge 1 &\forall \object{i}{u}, \object{i}{v} \in \objectSet{i}\label{pos0} \\
	& x_\object{i}{u} + w_\object{i}{u} \le x_\object{i}{v} + M \cdot (1 - r^{i,1}_{u,v}) &\forall \object{i}{u}, \object{i}{v} \in \objectSet{i}\label{pos1} \\ 
	& x_\object{i}{v} + w_\object{i}{v} \le x_\object{i}{u} + M \cdot (1 - r^{i,2}_{u,v}) &\forall \object{i}{u}, \object{i}{v} \in \objectSet{i} \\ & y_\object{i}{u} + h_\object{i}{u} \le y_\object{i}{v} + M\cdot  (1 - r^{i,3}_{u,v}) &\forall \object{i}{u}, \object{i}{v} \in \objectSet{i} \\ & y_\object{i}{v} + h_\object{i}{v} \le y_\object{i}{u} + M \cdot (1 - r^{i,4}_{u,v}) &\forall \object{i}{u}, \object{i}{v} \in \objectSet{i}\label{pos2} \\ 
    & w_\blockInstance{i}{k} = \width{\blockToInstance{i}{k}} & \forall \blockInstance{i}{k} \in \blockInstanceSet{i}\label{eq:blockw}\\
    &h_\blockInstance{i}{k} = \height{\blockToInstance{i}{k}}& \forall \blockInstance{i}{k} \in \blockInstanceSet{i}\label{eq:blockh}\\
    & \sum_{t=1}^{|\mathcal{D}^i_j|} s^i_{j,t} = 1& \forall \rectangle{i}{j} \in \rectangleSet{i}\label{eq:rectw}\\
	& w_\rectangle{i}{j} = \sum_{t=1}^{|\mathcal{D}^i_j|} w^i_{j,t} \cdot s^i_{j,t},~~h_\rectangle{i}{j} = \sum_{t=1}^{|\mathcal{D}^i_j|} h^i_{j,t} \cdot s^i_{j,t} & \forall \rectangle{i}{j} \in \rectangleSet{i}\label{eq:recth}\\
    & x_\object{i}{u}, y_\object{i}{u}, w_\object{i}{u}, h_\object{i}{u} \in \mathbb{R}_0^+  &\forall \object{i}{u} \in \objectSet{i} \\
    & \width{i},\height{i} \in \mathbb{R}_0^+ \\
    & r_{u,v}^{i,1},r_{u,v}^{i,2},r_{u,v}^{i,3},r_{u,v}^{i,4} \in \left\{0,1\right\} & \forall \object{i}{u}, \object{i}{v} \in \objectSet{i} \\
    & s_{j,t}^{i} \in \left\{0,1\right\} & \forall \rectangle{i}{j} \in \rectangleSet{i}~ \forall t \in \left\{1,\dots,|\mathcal{D}^i_j|\right\}     
\end{align}
\endgroup

The real variables $x,y,w,h$ model the positions and dimensions of the objects in the block type, while $\width{i},\height{i}$ model its boundary (\cref{eq:11,eq:12}). Non-overlapping is resolved using big-M constraints in \cref{pos0}-\cref{pos2}. There, binary variables $r^{i,k}_{u,v}$ determine whether $\object{i}{u}$ is to the left ($r^{i,1}_{u,v}=1$), right ($r^{i,2}_{u,v}=1$), below ($r^{i,3}_{u,v}=1$), or above ($r^{i,4}_{u,v}=1$) object $\object{i}{v}$. Finally, the dimensions of the objects need to be constrained. For rectangles, one of the available variants from set $\mathcal{D}^i_j$ is selected using binary variables $s^i_{j,t}$ in \cref{eq:rectw,eq:recth}, where $s^i_{j,t}=1$ means variant $t$ was selected for rectangle $\rectangle{i}{j}$. The size of block occurrences is coupled to the boundary variables of their relevant block type using \cref{eq:blockw,eq:blockh}. Note that these are the constraints that connect several partial single-block packing models into a monolithic 2DHRP model.

Additional valid constraints, which can be included, are those enforcing the ``absence of cycles''. Individually in left-right and up-down directions, topological ordering of the objects can be obtained with integer variables and the following constraints. A similar approach was used in \citet{topo}:
\begingroup
\allowdisplaybreaks
\begin{equation}
    g^{i,x}_u + 1 \le g^{i,x}_v + n \cdot (1 - r^{i,1}_{u,v}),~\forall \object{i}{u}, \object{i}{v} \in \objectSet{i}
\end{equation}
\begin{equation}
    g^{i,x}_v + 1 \le g^{i,x}_u + n \cdot (1 - r^{i,2}_{u,v}),~\forall \object{i}{u}, \object{i}{v} \in \objectSet{i}
\end{equation}
\begin{equation}
    g^{i,y}_u + 1 \le g^{i,y}_v + n \cdot (1 - r^{i,3}_{u,v}),~\forall \object{i}{u}, \object{i}{v} \in \objectSet{i}
\end{equation}
\begin{equation}
   g^{i,y}_v + 1 \le g^{i,y}_u + n \cdot (1 - r^{i,4}_{u,v}),~\forall \object{i}{u}, \object{i}{v} \in \objectSet{i}
\end{equation}
\begin{equation}
    g^{i,x}_u \in \left\{0,\dots,n \right\},~\forall \object{i}{u} \in \objectSet{i}\label{eq:endmilp}
\end{equation}
\endgroup

The values of the added variables $g^{i,x}_u$ correspond to the positions of objects in a topological ordering along the horizontal direction, while $g^{i,y}_u$ represent the corresponding ordering in the vertical direction. These constraints help eliminate cyclic dependencies in the relative positioning variables, thereby strengthening the formulation. In practice, their inclusion improves the performance of the MILP solver, while preserving all feasible solutions (with respect to object coordinates).

Altogether, we refer to ``partial MILP model'' consisting of equations \eqref{eq:11}-\eqref{eq:endmilp} (without any objective) for $\block{i}$ as $\textrm{Part}^{MILP}_\block{i}$. When partial models are combined across the hierarchy, we obtain a monolithic MILP model minimizing the half-perimeter of the top block type, further denoted as \texttt{M-MILP}:
\begingroup
\allowdisplaybreaks
\begin{align}
    \min \width{1} + \height{1}& \\
    \textrm{Part}^{MILP}_\block{i}&~~~\forall \block{i} \in \blockSet
\end{align}
\endgroup

\subsection{CP Model}\label{sec:cp}

We also provide a CP model as an alternative to the MILP model. Interval variables are used to model both position and dimensions of all rectangular objects. For each block type $\block{i}, i \in \left\{1,\dots,n\right\}$ the following model is created:
\begingroup
\allowdisplaybreaks
\begin{align}
    & \width{i} = \max_{\forall \object{i}{u} \in \objectSet{i}} \mathtt{endOf}(x_\object{i}{u}) \label{eq:cp1}\\
    & \height{i} = \max_{\forall \object{i}{u} \in \objectSet{i}} \mathtt{endOf}(y_\object{i}{u})\label{eq:cp2} \\
    &\texttt{endOf}(x_\object{i}{u}) \le \texttt{startOf}(x_\object{i}{v}) \vee \notag\\ & \texttt{endOf}(x_\object{i}{v}) \le \texttt{startOf}(x_\object{i}{u}) \vee \notag\\ & \texttt{endOf}(y_\object{i}{u}) \le \texttt{startOf}(y_\object{i}{v}) \vee \notag\\ & \texttt{endOf}(y_\object{i}{v}) \le \texttt{startOf}(y_\object{i}{u})& \forall \object{i}{u},\object{i}{v} \in \objectSet{i}\label{eq:cprel}\\    
    & \texttt{lengthOf}(x_\blockInstance{i}{k}) = \width{\blockToInstance{i}{k}}  & \forall \blockInstance{i}{k} \in \blockInstanceSet{i}\label{eq:cpblockw}\\
    & \texttt{lengthOf}(y_\blockInstance{i}{k}) = \height{\blockToInstance{i}{k}} & \forall \blockInstance{i}{k} \in \blockInstanceSet{i}\label{eq:cpblockh}\\
    & \mathtt{alternative}(x_\rectangle{i}{j}, [w^i_{j,1},\dots,w^i_{j,|\mathcal{D}^i_j|}]) & \forall \rectangle{i}{j} \in \rectangleSet{i} \label{eq:cprectw}\\
    & \mathtt{alternative}(y_\rectangle{i}{j}, [h^i_{j,1},\dots,h^i_{j,|\mathcal{D}^i_j|}]) & \forall \rectangle{i}{j} \in \rectangleSet{i} \\
    & \mathtt{presenceOf}(w^i_{j,t}) = \mathtt{presenceOf}(h^i_{j,t}) & \forall j \in \left\{1,\dots,m_i\right\}\label{eq:cprecth}\\&&\forall t \in \left\{1,\dots,|\mathcal{D}^i_j|\right\} \nonumber\\
    &x_\object{i}{u}~:~\mathtt{intervalVar}& \forall \object{i}{u} \in \objectSet{i} \\
    &y_\object{i}{u}~:~\mathtt{intervalVar}& \forall \object{i}{u} \in \objectSet{i} \\
    &w^i_{j,t}~:~\mathtt{optIntervalVar}& \forall j \in \left\{1,\dots,m_i\right\}\\&&\forall t \in \left\{1,\dots,|\mathcal{D}^i_j|\right\} \nonumber\\
    &h^i_{j,t}~:~\mathtt{optIntervalVar}& \forall j \in \left\{1,\dots,m_i\right\}\\&&\forall t \in \left\{1,\dots,|\mathcal{D}^i_j|\right\} \nonumber\\
    & \width{i}, \height{i}~:~ \mathtt{integerVar} 
\end{align}
\endgroup

The dimensions and positions of the objects are described by the properties of the interval variables $x,y$, and the boundary by the integer variables $W^i,H^i$. Boundary constraints are enforced by \cref{eq:cp1,eq:cp2}, and the absence of overlaps is achieved by \cref{eq:cprel}. Block occurrences are related to their relevant block types by \cref{eq:cpblockw,eq:cpblockh}. The selection of variants of the rectangles is done using optional interval variables in \cref{eq:cprectw}-\cref{eq:cprecth}. Note that the optional interval variables have fixed length (given the variant with which they are associated); the length of $x$ and $y$ intervals of rectangles is free, and the solver fixes them using the $\texttt{alternative}$ constraints.

The well-known concept of cumulative-resource constraints, powerful in project scheduling,  can also be advantageously used to add valid inequalities, thereby tightening the constraints. For each block type $\block{i}$:
\begin{equation}
    \sum_{j=1}^{m_i}\sum_{t = 1}^{|\mathcal{D}^i_j|} \texttt{pulse}(w^i_{j,t}, \texttt{lengthOf}(h^i_{j,t})) \le H^i\label{eq:cumul}
\end{equation}
\begin{equation}
    \sum_{j=1}^{m_i}\sum_{t = 1}^{|\mathcal{D}^i_j|}  \texttt{pulse}(h^i_{j,t}, \texttt{lengthOf}(w^i_{j,t})) \le W^i \label{eq:cumul2}
\end{equation}

\texttt{pulse}($where$,$height$) creates a signal, that is equal to $height$ where the interval $where$ is present, and 0 otherwise. These equations apply when there are no block occurrences within $\block{i}$. The single-dimensional cumulative constraints ensure that resource consumption (in case of \cref{eq:cumul}, the resource consumption refers to the length of the associated ``other-dimension'' interval) does not exceed capacity (total height for \cref{eq:cumul}) for any value of the respective $x$ or $y$ coordinate. The effect of block occurrence $\blockInstance{i}{k}$ can be included in the mentioned constraints using \texttt{heightOf} operator, which passes the dynamic width and height $\width{\blockToInstance{i}{k}},\height{\blockToInstance{i}{k}}$ to newly constructed pulses. 

Altogether, equations \eqref{eq:cp1}-\eqref{eq:cumul2} form the ``partial CP model'' $\textrm{Part}_\block{i}^{CP}$ for block type $\block{i}$. The monolithic model $\texttt{M-CP}$ is obtained as:
\begingroup
\allowdisplaybreaks
\begin{align}
    \min  \width{1}+ \height{1}& \\
    \textrm{Part}^{CP}_\block{i}&~~~\forall \block{i} \in \blockSet
\end{align}
\endgroup

%% file: sections/graph.tex
\resizebox{\textwidth}{!}{%
\begin{tikzpicture}[
  >={Stealth},
  vertex/.style={circle, draw=black, font=\sffamily\bfseries},
  purple/.style={fill={rgb,255:red,128; green,0; blue,128}, fill opacity=0.65},
  blue/.style={fill={rgb,255:red,30; green,144; blue,255}, fill opacity=0.65},
  green/.style={fill={rgb,255:red,0; green,128; blue,0}, fill opacity=0.65},
  red/.style={fill={rgb,255:red,255; green,0; blue,0}, fill opacity=0.65},
  level distance=15mm
]

\Large
\node[vertex,purple,minimum size=19mm] (A) {$\block{1}$};

\node[vertex,blue,minimum size=19mm,below left=of A] (B) {$\block{2}$};
\node[vertex,green,minimum size=19mm,right=of B] (C) {$\block{3}$};
\node[vertex,purple,minimum size=8mm,right=of C](a) {$\rectangle{1}{1}$};
\node[vertex,purple,minimum size=8mm,right=of a] (b) {$\rectangle{1}{2}$};
\node[vertex,purple,minimum size=8mm,right=of b] (c) {$\rectangle{1}{3}$};

\node[vertex,red,minimum size=19mm,below left=of B] (D) {$\block{4}$};
\node[vertex,blue,minimum size=8mm,right=of D] (d) {$\rectangle{2}{1}$};
\node[vertex,green,minimum size=8mm,right=of d] (e) {$\rectangle{3}{1}$};
\node[vertex,green,minimum size=8mm,right=of e] (f) {$\rectangle{3}{2}$};

\node[vertex,red,minimum size=8mm,below=of D] (g) {$\rectangle{4}{1}$};
\node[vertex,red,minimum size=8mm,right=of g] (h) {$\rectangle{4}{2}$};

\graph {
  (A) ->[edge label=2] (B);
  (A) ->[edge label=1, at end] (C);
  (A) ->[edge label=1, near end] (a);
  (A) ->[edge label=1, near end] (b);
  (A) ->[edge label=1, near end] (c);

  (B) ->[edge label=1] (D);
  (B) ->[edge label=1, near end] (d);

  (C) ->[edge label=1, near end] (e);
  (C) ->[edge label=1, near end] (f);

  (D) ->[edge label=1, near end] (g);
  (D) ->[edge label=1, near end] (h);
};
\end{tikzpicture}
}

%% file: sections/4-decomposition.tex
\section{Baseline Decomposition Methods}\label{sec:baselines}
\subsection{Heuristics and Lower Bounds}\label{sec:heuristic}
Due to the complexity of the single-block packing problem alone, it is necessary to provide good initial solutions to the MILP or CP solvers. Two well-known heuristics can be used: the bottom left heuristic \citep{bottomleftfill} and the best fit heuristic \citep{bestfit}. The bottom left heuristic utilizes the simplified approach of \citet{martello-minimum-square}: objects are packed one-by-one in the bottom left manner, using a single permutation of objects sorted by their area. 

The best fit heuristic is run three times, to incorporate all three position selection strategies described by \citet{bestfit}. Namely, whenever the best fit heuristic determines a segment and a rectangle to place, it is positioned according to one of the following rules:  (a) at the leftmost position within the segment, (b) adjacent to the taller of the rectangles defining the segment, or (c) adjacent to the shorter of the rectangles defining the segment boundary. The heuristic expects one of the dimensions of the block type to be fixed. If there is no such constraint (e.g., for the top block type minimizing half-perimeter), the width of the block type is set to the square root of the block type's expected area, estimated from its rectangles. 

The runtime of both heuristics is negligible, and both of them are called whenever a solution to a single-block packing problem is needed; whenever a CP or MILP solver is to be started, and there is no existing solution, heuristics provide a warm start. Furthermore, solutions of lower-level block types can be utilized to construct solutions for upper-level block types, constructing an initial solution for the entire monolithic model rapidly.

Due to the presence of variants, the lower bounds developed for strip packing in, e.g., \citet{alvarez-lb}, cannot be directly used. Thus, the minimum area bounds are calculated in the following manner. For each block type $\block{i}$, area estimates across its block occurrences $\blockInstanceSet{i}$ and areas of rectangles $\rectangleSet{i}$ (specifically, their smallest variant) are combined: 
\begin{equation}\label{eq:area-estimate}
    \LBarea{\block{i}} =  \sum_{\blockInstance{i}{k} \in \blockInstanceSet{i}} \LBarea{\block{\blockToInstance{i}{k}}}+\sum_{j=1}^{m_i} \min_{t \in \left\{1,\dots,|\mathcal{D}^i_j|\right\}} w^i_{j,t}\cdot h^i_{j,t}
\end{equation}

The main point of interest is the lower bound of the top block type, denoted $\LBarea{\block{1}}$, which we abbreviate as $\LBareatop$. Based on this bound, a corresponding lower bound on the minimum half-perimeter bound can also be derived:
\begin{equation}
    \LBhp{\block{i}} = 2\cdot \sqrt{\LBarea{\block{i}}}
\end{equation}

The half-perimeter bound for the top block type $\LBhp{\block{1}}$ is shortened to $\LBhptop$.

\subsection{Bottom-Up Decomposition Method}\label{sec:bottom-up}
Even with a good initial solution, solvers using monolithic models outlined in \cref{sec:desc} struggle to optimize large instances within the limited computation time. Therefore, it is natural to decompose the problem and solve it in parts, even without an optimality guarantee. A simple way to do this is to use the Bottom-Up method, which was utilized in \citet{xu-placement,zhu-circuits}. Due to the out-tree hierarchy, packing solutions for the leaf block types can be constructed directly, and these solutions are subsequently provided to the upper-level block types. It is only necessary to pass information about the width and height of the child block types, whose block occurrences behave as rectangles with a set of newly generated variants. When all block types are processed (in the reverse topological order), a feasible solution of the original problem is obtained.

At the top level, the half-perimeter of the block type is minimized. At lower levels, however, the goal is to produce packing solutions that can integrate well with other block occurrences and rectangles at higher levels of the hierarchy. Since the hierarchy is traversed in a bottom-up manner, no information from upper levels is available when solving lower-level subproblems. To mitigate potential incompatibilities between block type dimensions (e.g., one being very wide and the other being very tall), multiple packing variants are generated for each block type. This provides a diverse set of candidate shapes, increasing the likelihood that compatible combinations can be formed at higher levels. This is achieved by generating a set of maximum widths $\width{i}_{\max} = \left\{\width{i}_{\max, 1}, \dots, \width{i}_{\max, N}\right\}$ for each block type and solving the single-block strip-packing problem for each such width. The number of possible packing variants for each block type depends on the desired number of variants $N$. Generating more variants increases the likelihood that at least one of them will integrate well with other objects at higher levels of the hierarchy. However, in practice, the overall computational budget is limited. Since variants are optimized sequentially, increasing their number reduces the computational effort that can be devoted to each individual variant under a fixed time budget. Therefore, the number of variants per block type must be carefully controlled in order to balance solution diversity and optimization quality.

In addition, it is crucial to make a good selection of the $\width{i}_{\max}$ values. In this paper, they are selected by uniformly partitioning the suitable range of aspect ratios (widest and tallest possible packing), and calculating the dimensions given area estimate $\LBarea{\block{i}}$. Altogether, a feasible packing variant with maximum width $\width{i}_{\max, q}$ is obtained by solving the following packing problem with a strip-packing-like objective:
\begin{equation}
    \min \height{i},~s.t.~ \width{i} \le \width{i}_{\max,q} \wedge \eqref{eq:a1}-\eqref{eq:a3} 
\end{equation}
 
This is done for each block type and each variant $q$. The no-overlap and boundary constraints \cref{eq:a1,eq:a2,eq:a3} are enforced by the relevant MILP and CP constraints of \cref{sec:milp,sec:cp} for the chosen formalism. Note that there are no block occurrences in such a single-block model, since they were replaced by packing variants from the child block types; yet, each ``block occurrence rectangle'' of the same child block type $\block{i'}$ still has to use the same identical variant.  

Since our objective is to obtain high-quality solutions within a limited computational budget, we assume that a fixed total time $T$ is available for solving the entire instance. This budget must be distributed among all block types. One possible allocation strategy, which we adopt in this paper, is to distribute time proportionally to the number of objects. Specifically, the computation time $\tau(\block{i})$ allocated to $\block{i}$ is defined as:
\begin{equation}
    \tau(\block{i}) = \frac{|\blockInstanceSet{i}| + |\rectangleSet{i}|}{\sum_{\forall\block{j}\in\blockSet} |\blockInstanceSet{j}| + |\rectangleSet{j}|} \cdot T
\end{equation}

This heuristic allocates more time to block types containing a larger number of objects, which are expected to be more computationally demanding. Such an approach has also been used in the context of integrated circuit placement \citep{xu-placement}. The experimental results reported in \cref{exp:single-level} show that, for both MILP and CP solvers, solution quality deteriorates as the number of rectangles increases when the time limit is fixed. Finally, the allocated time $\tau(\block{i})$ is uniformly divided among the packing variants of $\block{i}$ described earlier. We emphasize that this proportional allocation is a design choice, and alternative strategies could also be considered.

Overall, this Bottom-Up approach is referred to as $\texttt{BU}_N$, where $N$ is the number of packing variants generated for each block type with the exception of the top block type. Note that more complex time management and variant generation strategies could be employed. Finally, any solver could be used to solve the isolated single-block packing problem that is encountered for each block type and packing variant. We elaborate on the choice of the solver in \cref{sec:experiments}. This also means that the solution provided by heuristics in \cref{sec:heuristic} could be directly used as a feasible packing variant, without explicitly utilizing any MILP or CP solver. This fully heuristic variant of the Bottom-Up approach is denoted as \texttt{HEUR} in \cref{sec:experiments}. Note that the solution obtained with $\texttt{HEUR}$ is used to warm-start monolithic models of \cref{sec:desc}.

 The Bottom-Up works very well despite its simplicity. However, it cannot reason which packing variants would be useful at the top node. To counter this weakness, Bottom-Up needs to generate multiple variants to ensure one of them actually works well. This wastes computation time (by investigating useless variants) and requires additional control by the user (how many/which variants, time management, etc.). This motivated us to develop a more informed decomposition method, which could outperform the Bottom-Up baseline.

%% file: sections/4b-decomposition.tex
\section{Logic-based Benders Decomposition-like Method}\label{sec:benders}

In this section, we describe the main contribution of our paper: the LBBD-based method, which aims to overcome the drawbacks of the Bottom-Up method mentioned in the last paragraph of \cref{sec:bottom-up}. We first describe the decomposition in its exact formulation, enabling us to find an optimal solution. Later, in \cref{sec:proposed-mod}, a heuristic version of this decomposition is described that produces good solutions in a limited time. That version is used in the majority of experiments. 

\subsection{Decomposition Scheme}\label{sec:summary}

The decomposition method solves the 2DHRP considering one block type at a time. The rest of this paragraph summarizes the high-level overview. The method starts from the top-level block type and formulates a ``master problem'' in which lower-level block occurrences are represented using only coarse information, namely their estimated areas (instead of working with explicit packing variants as the Bottom-Up method in \cref{sec:bottom-up}). Once an optimal solution for this ``master problem'' is found, child subproblems are invoked for each child block type to verify whether the corresponding block occurrences can be feasibly packed within the assigned dimensions. If all child subproblems are feasible, the solution is optimal, as discussed in \cref{sec:proof}. Otherwise, cuts are generated within the ``master problem''. They eliminate the previously suggested sizes of block occurrences and refine the representation of the corresponding child block types. The process then iterates. Due to the recursive nature of the problem, if a child subproblem itself contains other block types, the same procedure is applied recursively. Consequently, the notion of a ``master problem'' is relative to the currently considered block type and is dynamically reassigned as the algorithm explores the hierarchy.

\begin{figure}[htbp]
    \centering
    
        \begin{subfigure}[b]{0.49\textwidth}  
            \centering 
            \includegraphics[width=0.95\textwidth]{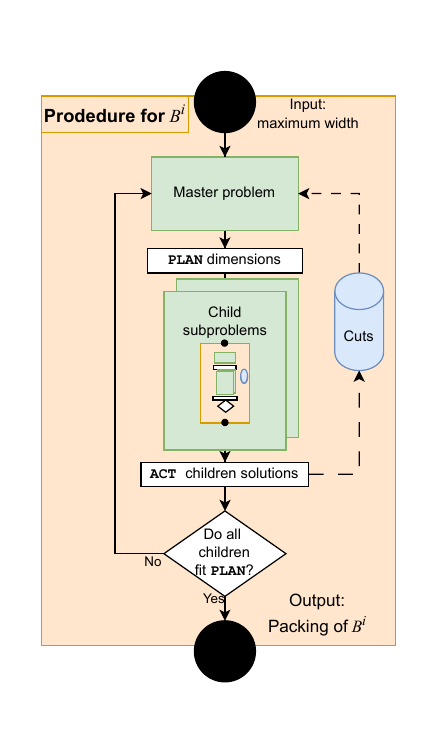}
            \caption{Diagram of the recursive procedure of the LBBD for block type $\block{i}$.} 
            \label{fig:diagram-small}
        \end{subfigure}
        \hfill
        \begin{subfigure}[b]{0.49\textwidth}
            \centering
            \includegraphics[width=0.99\textwidth]{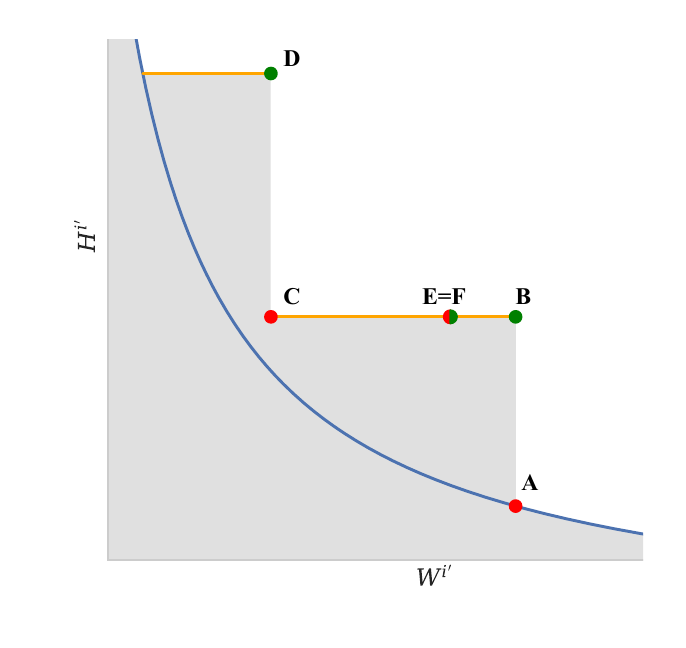}
            \caption{Solution space of the child block type $\block{i'}$ in the ``master problem'' model of $\block{i}$. Initial hyperbola and two cuts dividing the white region of potentially feasible pairs $\width{i'},\height{i'}$, $i'\in\mathcal{C}^i$ from the infeasible pairs. Red dots A, C, E are examples of \texttt{PLAN} pairs provided by $\block{i}$ in 1st, 2nd, and 3rd iteration, while green dots B, D, F are subsequent dimensions \texttt{ACT} of $\block{i'}$.}
            \label{fig:contour-small}
        \end{subfigure}
        \caption{Illustrative diagrams describing how the proposed LBBD method works.} 
        \label{fig:method-small}
\end{figure}

This is shown in a recursive algorithm as illustrated in \cref{fig:method-small}. \cref{fig:diagram-small} shows the control flow diagram of the procedure for block type $\block{i}$. The input of the procedure is an additional constraint (none for the top block type $\block{1}$, and maximum allowed width for any other block type), and the output of the procedure is the packing for $\block{i}$ that respects all constraints, and minimizes either half-perimeter, or just height. The algorithm starts by processing the top block type, where it tries to minimize its half-perimeter, 
 and recursively enters the respective block type's children with a maximum width constraint derived from the parent, while minimizing the height of the block type.

At each $\block{i}$, the ``master problem'' (which only approximates the dimensions of child block types' occurrences) needs to be solved. Only the partial model (\eqref{eq:11}-\eqref{eq:endmilp}  for MILP) of $\block{i}$ from \cref{sec:desc} is constructed, and the additional constraint passed from the parent block type of $\block{i}$ is added. Without the partial models for the children of $\block{i}$, the width and height variables $\width{i'},\height{i'}$, $i' \in \mathcal{C}^i$ (indices of children of $\block{i}$) would be free variables, and the dimensions of the block occurrences would be set arbitrarily (see \cref{eq:blockw,eq:blockh,eq:cpblockw,eq:cpblockh}). Therefore, these variables need to be constrained in another manner. This is done by using the constant area bound:
\begin{equation}
    \width{{i'}} \cdot \height{{i'}} \ge \LBarea{\block{i'}},~\forall i' \in \mathcal{C}^i\label{eq:area-estimated-eq}
\end{equation}

In the MILP, this is approximated by a piecewise linear function. The set of potentially feasible solutions given by this constraint for $\block{i'}, i' \in\mathcal{C}^i$ is the convex region delimited by the blue hyperbola shown in \cref{fig:contour-small}. Thus, the dimensions of occurrences of $\block{i'}$ at least follow the smallest possible area. These constraints, partial model from \cref{sec:desc} (ensuring objects fitting the boundary and their non-overlapping), and additional cuts (described later) form the ``master problem'' in \cref{fig:diagram-small}. For MILP, this model for some (non-top) block type $i$ is:
\begin{align}
    &\min \height{i}&\label{eq:master1} \\
    & \width{i} \le \width{i}_{\texttt{PLAN}} \label{eq:parent-constraint} \\
    &\textrm{Part}^{MILP}_\block{i} \label{eq:master-partial}\\
    &\width{{i'}} \cdot \height{{i'}} \ge \LBarea{\block{i'}}&\forall i' \in  \mathcal{C}^i\\
    &cuts\label{eq:master2} 
\end{align}

The solution to this problem is depicted as $\texttt{PLAN}$ in \cref{fig:diagram-small} and is influenced by constraint \cref{eq:parent-constraint} from the parent of $\block{i}$. However, it may not be a feasible partial solution for the original 2DHRP problem. It is necessary to verify that the dimensions of each child block type $\block{i'}, i'  \in \mathcal{C}^i$ are valid; i.e., the child block type can be truly packed into a boundary with dimensions $\width{i'}_\texttt{PLAN}, \height{i'}_\texttt{PLAN}$, which are the dimensions of block occurrences of $\block{i'}$ in the ``master problem'' solution.

For that purpose, the same procedure of \cref{fig:diagram-small} is started for each $\block{i'}$. $\block{i'}$ again optimizes model \eqref{eq:master1}-\eqref{eq:master2}: minimizing $H^{i'}$ and adding the constraint $W^{i'} \le W^{i'}_\texttt{PLAN}$ found in $\block{i}$. The procedure is recursively initiated for possible children of $\block{i'}$.

Eventually, a feasible packing of $\block{i'}$ satisfying the imposed constraints is obtained (since there always exists at least one feasible packing, e.g., one rectangle on top of another). This packing solution, denoted as solution \texttt{ACT} (see \cref{fig:diagram-small}), has actual feasible dimensions $\width{i'}_\texttt{ACT},\height{i'}_\texttt{ACT}$. Solution \texttt{ACT} is returned to $\block{i}$.

The algorithm for $\block{i}$ checks whether the actual packing of $\block{i'}$ follows the suggested dimensions. If $\height{i'}_\texttt{PLAN} \ge \height{i'}_\texttt{ACT}~\forall i' \in \mathcal{C}^i$, the solution of the ``master problem'' of $\block{i}$ is feasible with respect to the children subproblems, and optimal packing of the original problem has been found. Otherwise, the following cut is added to the ``master problem'' model to reduce the space of potentially feasible dimensions for each child block type $\block{i'}$ that could not be verified:
\begin{equation}\label{eq:cut}
    \width{i'} \le \width{i'}_\texttt{PLAN} \implies \height{i'} \ge \height{i'}_\texttt{ACT}
\end{equation}

If no solution was found, $\width{i'}_\texttt{PLAN}$ was too narrow, and we add a cut:
\begin{equation}\label{eq:cut-small}
    \width{i'} > \width{i'}_\texttt{PLAN}
\end{equation}
Then, the ``master problem'' for $\block{i}$ is solved again with these additional cuts. If it is solved optimally, these cuts remove only infeasible pairs $\width{i'},\height{i'}$, and after a sufficient number of iterations of the loop of \cref{fig:diagram-small}, the ``master problem'' would result in a solution that is feasible given the children of $\block{i}$. This solution is an optimal solution to the original problem.

The way in which cuts \eqref{eq:cut} reduce the search space is shown in \cref{fig:contour-small}. There, potentially feasible pairs of width and height of one of the child block type $\block{i'},~i'\in\mathcal{C}^i$, correspond to the white region, while infeasible pairs correspond to the gray region. Potentially feasible pairs refer to dimensions of block occurrences of $\block{i'}$, that are feasible with respect to the current set of cuts introduced for $\block{i'}$. With each iteration of \cref{fig:diagram-small}, a new cut derived from the verification of a new pair $\width{i'}_\texttt{PLAN},\height{i'}_\texttt{PLAN}$, may be added to model $\block{i'}$ more precisely.

The first reduction of potentially feasible pairs is done by the hyperbolic curve corresponding to \cref{eq:area-estimated-eq}. In the first iteration, \texttt{PLAN} dimension pair (A) was suggested as the ``master problem'' solution of $\block{i}$, and taller \texttt{ACT} dimension pair (B) was verified as the $\block{i'}$ subproblem. This generated the first cut.

In the next iteration, a new \texttt{PLAN} dimension pair (C) was suggested, but again the child solution had a greater height (D). Finally, \texttt{PLAN} (E) was successfully verified in the third iteration by (F). The cuts obtained by the first two iterations reduced the space by introducing two ``stairs''. Note that the orange tops (including B, D, E/F) of the stairs belong to a potentially feasible region, while the vertical faces (including A and C) do not.

\subsection{Convergence Proof}\label{sec:proof}

In this section, we prove that the proposed LBBD finds an optimal solution in a finite number of steps. The proof is provided for an instance with two levels of hierarchy and two block types $\block{1}$ and $\block{2}$; however, it can be generalized for an arbitrary number of levels, since solving the child subproblem for any intermediate block type with children of its own performs the same procedure recursively. We assume that any single-level packing problem is solved optimally (by MILP or CP solver).

\begin{lemma}
The LBBD algorithm converges to an optimal solution of the 2DHRP problem in a finite number of iterations.
\end{lemma}
\begin{proof}
The ``master problem`` optimizing the half-perimeter of the top block type $\block{1}$ contains only bounded integer variables, hence its feasible region is finite. At each iteration, the child subproblem checks whether $\block{2}$ can be packed within the provided width $W^2_\texttt{PLAN}$ and height $H^2_\texttt{PLAN}$. If a solution was found with $W^2_\texttt{ACT}\leq  W^2_\texttt{PLAN}$ and  $H^2_\texttt{ACT} >  H^2_\texttt{PLAN}$, then a valid cut \eqref{eq:cut} is generated and added to the ``master problem''. This cut removes the previous ``master problem'' solution, since block occurrences of $\block{2}$ with dimensions suggesting $W^2 \leq W^2_\texttt{PLAN},~H^2 \leq H^2_\texttt{PLAN} < H^2_\texttt{ACT}$ can no longer be proposed. Therefore, no solution can be generated more than once. Otherwise, if no solution was found, the cut \eqref{eq:cut-small} operates in a similar manner. Since the number of feasible solutions is finite, the algorithm terminates after a finite number of iterations.

Cuts \eqref{eq:cut} do not remove any feasible solution to the original problem. They do not affect the feasibility region of the ``master problem'' for $W^2 > W^2_\texttt{PLAN}$. For $W^2 \le W^2_\texttt{PLAN}$, no feasible solution exists with $H^2 < H^2_\texttt{ACT}$. We have assumed that the optimal solution of the child subproblem was found. However, if there was packing of $\block{2}$ with $W^2 \le W^2_\texttt{PLAN}, H^2 < H^2_\texttt{ACT}$, such a packing would be a feasible and strictly better solution of the child subproblem, which is a contradiction. Cuts \eqref{eq:cut-small} do not remove any feasible solution, since they are generated when there is no feasible solution for given $W^2_\texttt{PLAN}$.

Upon termination, all subproblems are feasible, and no cut can be added to prune the found solution. Therefore, the solution of the ``master problem'' is feasible for the original problem. Since the ``master problem'' is solved to optimality at each iteration, the final solution is optimal.
\end{proof}

\subsection{Heuristic-LBBD}\label{sec:proposed-mod}

\begin{figure}[htbp]
    \centering
    
        \begin{subfigure}[b]{0.49\textwidth}  
            \centering 
            \includegraphics[width=0.95\textwidth]{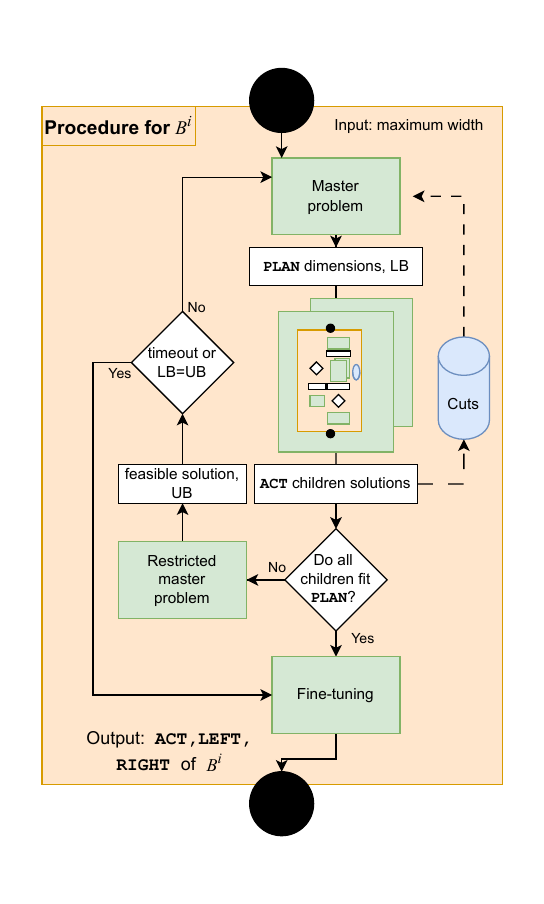}
            \caption{Diagram of the recursive procedure of the heuristic-LBBD for block type $\block{i}$.} 
            \label{fig:diagram}
        \end{subfigure}
        \hfill
        \begin{subfigure}[b]{0.49\textwidth}
            \centering
            \includegraphics[width=\textwidth]{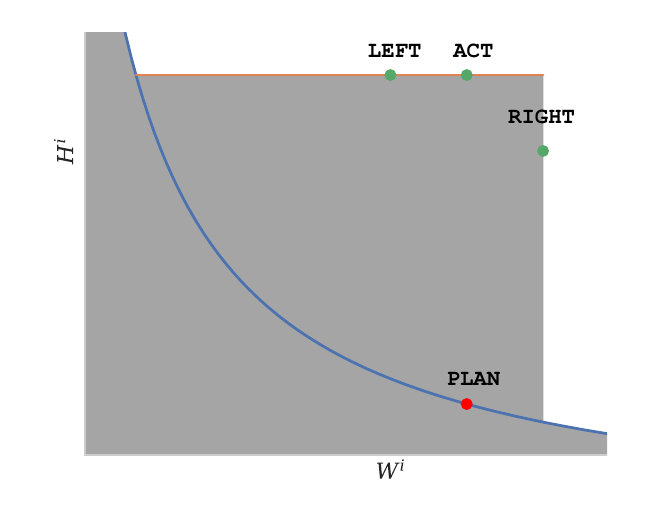}
            \caption{Solution space of the current block type $\block{i}$ in the ``master problem'' model of its parent. Example cut and width-height pairs provided to the parent of $\block{i}$ during fine-tuning of $\block{i}$.}
            \label{fig:contour}
        \end{subfigure}
        \caption{Illustrative diagrams describing how the implemented heuristic-LBBD operates.} 
        \label{fig:method}
\end{figure}

The method described in \cref{sec:summary} produces an optimal solution, but it relies on an optimal solution being found for each single-block packing problem to produce valid cuts, which is time-consuming and, even for small-sized instances, makes the method practically inapplicable. In this section, several heuristic modifications are developed, sacrificing optimality of the method but achieving good results in a reasonable time. The recursive procedure works in a similar way as in \cref{sec:summary}, but its control flow diagram is extended to stop optimization early, as shown in \cref{fig:diagram}.

\subsubsection{Limited Computation Time and Solutions without Proven Optimality}

First, while the original LBBD automatically decides which dimensions of block types to explore, we still need to manage the allocation of the computation time. While the ``master problem'' is optimized, periodic checks are performed to determine whether the solver has found a new solution. If no improvement was achieved within the local time limit (improvement period), computation is aborted, and the current (possibly non-optimal) solution $\texttt{PLAN}$ is returned. 

Note that such a solution, when passed from child block type to parent block type, could generate a cut (as in \cref{fig:contour}) that may remove some otherwise feasible width-and-height pairs. This means that the cut overconstrains the problem as the dark gray area no longer contains only infeasible pairs, as was the case in \cref{sec:summary}, but may also contain feasible pairs. That could prevent an optimal solution from being found, since the solver cannot utilize them. However, such non-optimal approaches often yield a good time-performance ratio \citep{heur-benders-1}. For the purpose of the heuristic-LBBD, we view the cuts as being valid, and a dimension pair being considered infeasible as a statement regarding this heuristic setting, not the original 2DHRP.

Since it may take many iterations of the original loop of \cref{fig:diagram-small} for the ``master problem'' to produce a feasible packing, we need to ensure that at least some packing is always found early. This is done by the ``restricted master problem'' step in \cref{fig:diagram}. This step solves the structurally similar model as in the ``master problem''. However, instead of constraining the width and height of the child block types to their area and the generated cuts, they are fixed to the dimensions of feasible packing found by the child block types: 
\begin{align}
    &\min \height{i}& \label{eq:restricted1} \\
    & \width{i} \le \width{i}_{\texttt{PLAN}} \\
    &\textrm{Part}^{MILP}_\block{i} \\
    &\width{i'}=\width{i'}_\texttt{ACT},~ \height{i'}=\height{i'}_{\texttt{ACT}}&\forall i' \in  \mathcal{C}^i\label{eq:restricted2} 
\end{align}

The solver is partially initialized with relative positions from the solution of the ``master problem''. Solution of the ``restricted master problem'' is actually a feasible packing of $\block{i}$ since it uses the feasible packing for each child. It is then used as an upper bound while iterating, and when the time limit allocated for the block type is reached, the best solution found so far is returned. 

\subsubsection{Fine-tuning}\label{sec:finetune}

When $\block{i}$ is being optimized, once the control flow leaves the main loop and enters ``Fine-tuning'' in \cref{fig:diagram}, there is an opportunity to improve the packing so the parent of $\block{i}$ can strengthen its cuts. From perspective of parent of $\block{i}$, let $\width{i}_\texttt{PLAN}$ be the width suggested by the parent, and let $\width{i}_\texttt{ACT},\height{i}_\texttt{ACT}$ be the dimensions of the best solution found after leaving the main loop in \cref{fig:diagram} for current block type $\block{i}$.

\textbf{Improving Width of \texttt{ACT}:} ~As in \cref{fig:diagram-small}, solution \texttt{ACT} is a feasible packing of $\block{i}$ (from one of the iterations of the loop). Since $\block{i}$ was optimized in a strip packing manner by minimizing the height, it can be further improved by minimizing its width. This is done by re-solving the same ``restricted master problem'' model, but with objective $\min \width{i}$ and setting $\height{i}\le \height{i}_\texttt{ACT}$. 
The solver is warm-started with the existing solution $\texttt{ACT}$. Once optimized, solution $\texttt{LEFT}$ is obtained, with $\width{i}_\texttt{LEFT}\le \width{i}_\texttt{ACT},\height{i}_\texttt{LEFT}\le \height{i}_\texttt{ACT}$. The solution should be ``to the left'' of \texttt{ACT} in \cref{fig:contour}, and thanks to the smaller width, it is an improvement on the original solution $\texttt{ACT}$. Both \texttt{LEFT} and \texttt{ACT} solutions are returned to the parent of $\block{i}$ once the control flow exits the diagram \cref{fig:diagram}. Then, the parent of block type $\block{i}$ adds the same cut as in \cref{eq:cut}:

\begin{equation}
    \width{i} \le \width{i}_\texttt{PLAN} \implies \height{i} \ge \height{i}_\texttt{LEFT}
\end{equation}

But if in the next iteration parent of block type $\block{i}$ suggests a new $\texttt{PLAN}^{\texttt{NEW}}$ dimensions with $\width{i}_\texttt{LEFT} \le \width{i}_{\texttt{PLAN}^{\texttt{NEW}}} \le \width{i}_\texttt{ACT}$, then this does not need to be validated by solving the subproblem for $\block{i}$ since a feasible child solution is already known. 


\textbf{Strengthening the Cut:}~Similarly, while still in the ``Fine-tuning'' part of the diagram for $\block{i}$, we can try to find a closest packing with a smaller height. To do this, the height decrease $\alpha$ is selected and the modified ``master problem'' model (with children of $\block{i}$ modeled using generated cuts) is solved:
\begin{equation}
    \min \width{i},~s.t.~\height{i}\le\height{i}_\texttt{ACT} - \alpha ~\wedge~ \eqref{eq:master-partial}-\eqref{eq:master2}
\end{equation}

Solution should satisfy: $\width{i}_\texttt{RIGHT} > \width{i}_\texttt{ACT}, \height{i}_\texttt{RIGHT} \le \height{i}_\texttt{ACT} - \alpha$. Due to the imprecise modeling of child block types, this solution may not be a feasible packing of $\block{i}$, but its width can be interpreted as a lower bound for fixed height $\height{i}_\texttt{ACT}-\alpha$ of $\block{i}$. Thus, once $\block{i}$ finishes and returns control to its parent, its parent adds a wider cut to better model $\block{i}$:
\begin{equation}
    \width{i} < \width{i}_\texttt{RIGHT} \implies \height{i} \ge \height{i}_\texttt{LEFT}
\end{equation}

This expands the original cut to the right, as is shown in \cref{fig:contour}; the parent of $\block{i}$ reduces the set of potentially feasible dimensions of its representation of $\block{i}$. If $\alpha=1$, the cut should not overestimate the height of any solution with width between $\width{i}_\texttt{ACT}$ and $\width{i}_\texttt{RIGHT}$. Risking this guarantee, a greater reduction of search space can be obtained by setting $\alpha$ to larger values, e.g., $\alpha=\lfloor 0.05\cdot \height{i}_\texttt{ACT}\rfloor$. Finally, the case with $\alpha=0$  corresponds to omitting the computation of $\texttt{RIGHT}$ altogether.

The result of fine-tuning can be seen in \cref{fig:contour}. We can see the initial hyperbola and an additional cut. Calculation was initiated by red ($\width{i}_\texttt{PLAN},\height{i}_\texttt{PLAN}$) pair from the parent of block type $\block{i}$. However, found ($\width{i}_\texttt{ACT},\height{i}_\texttt{ACT}$) has height greater than the original suggestion. Then, fine-tuning was performed and produced solutions ($\width{i}_\texttt{LEFT},\height{i}_\texttt{LEFT}$) and ($\width{i}_\texttt{RIGHT},\height{i}_\texttt{RIGHT}$). \texttt{LEFT} solution improved the existing \texttt{ACT} solution, \texttt{RIGHT} expanded the cut and significantly reduced the solution space, as the dark gray area suggests.

In \cref{sec:experiments}, we test three versions of our proposed method, abbreviated as \texttt{LBBD}: $\texttt{LBBD}_\texttt{0}$ does not use the decremented-height part of fine-tuning at all, $\texttt{LBBD}_\texttt{1}$ sets $\alpha=1$, and $\texttt{LBBD}_\texttt{R}$ uses the radical strategy with $\alpha=\lfloor 0.05\cdot \height{i}_\texttt{ACT}\rfloor$.

\subsection{Runtime Experiment and Recursive Procedure Illustration}

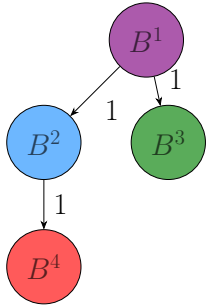
\begin{figure}
    \centering
    \input{sections/graph2}
    \caption{Illustrative instance's hierarchy with four block types. For simplicity, nodes for individual rectangles were omitted.}
    \label{fig:hie-eaxmple}
\end{figure}

\begin{figure}[htbp]
    \centering
    
        \begin{subfigure}[b]{0.49\textwidth}  
            \centering 
            \includegraphics[width=\textwidth]{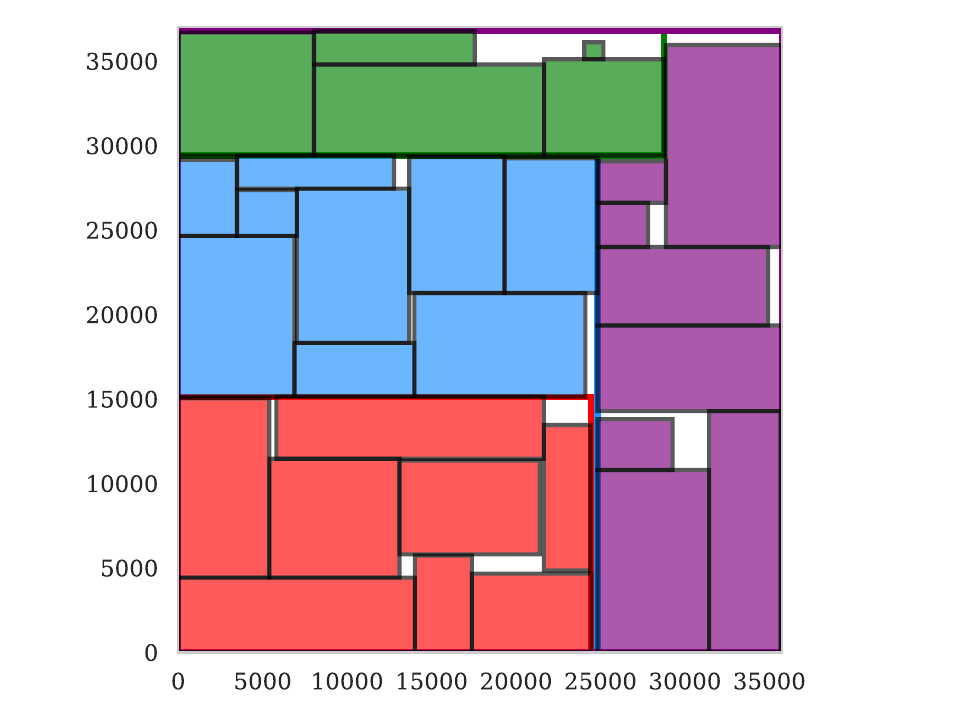}
            \caption{Solution obtained at 30 seconds, \newline$W+H=72509$.} 
            \label{fig:firstol}
        \end{subfigure}
        \hfill
        \begin{subfigure}[b]{0.49\textwidth}
            \centering
            \includegraphics[width=\textwidth]{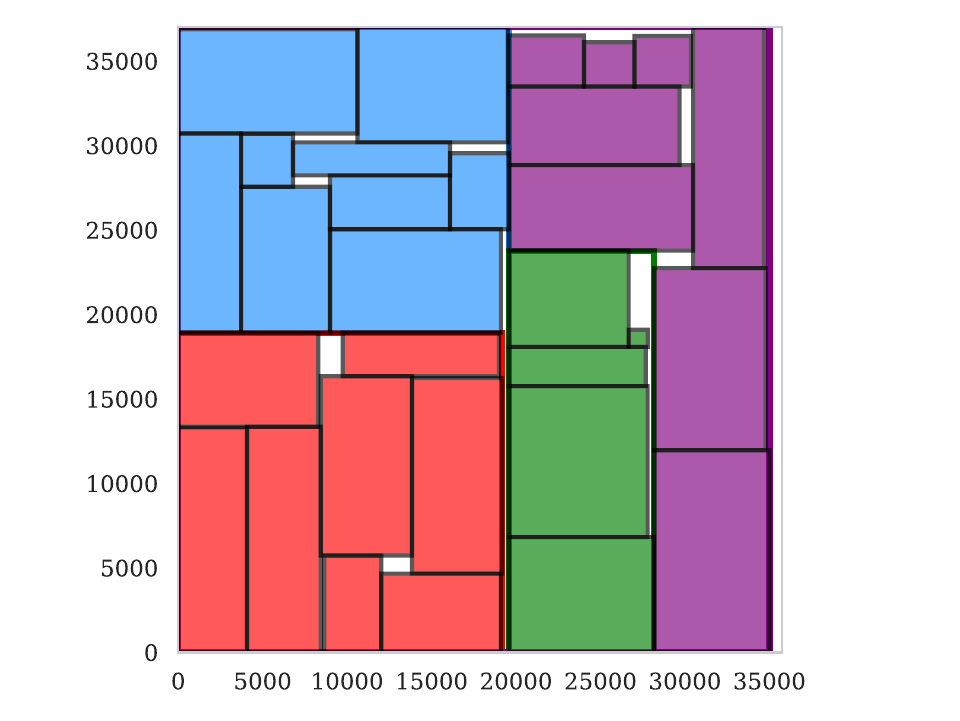}
            \caption{Solution obtained at 120 seconds,\newline $W+H=72019$.}
            \label{fig:secondsol}
        \end{subfigure}
        \caption{Initial (a) and first improving (b) solutions found by the heuristic-LBBD in its example run.} 
        \label{fig:firstsecond}
\end{figure}

In this section, the computation on the $\texttt{LBBD}_{1}$ is demonstrated using a three-level instance with hierarchy shown in \cref{fig:hie-eaxmple} with individual rectangles shown later in \cref{fig:firstsecond}.

\begin{figure}
    \centering
\includegraphics[width=0.98\linewidth]{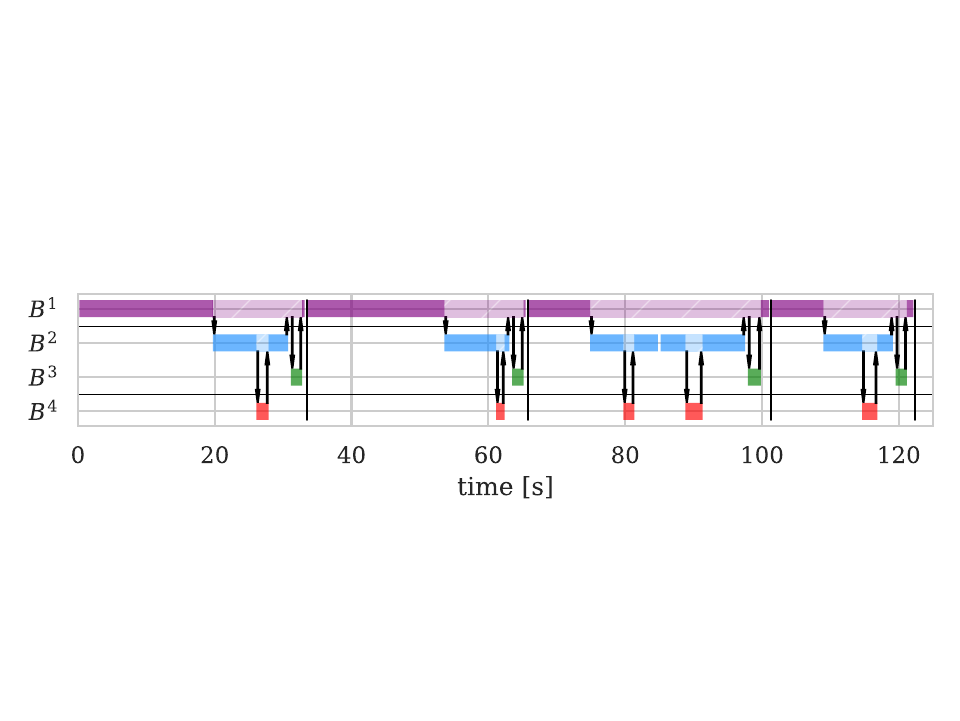}
    \caption{Interaction between individual block types in \cref{fig:hie-eaxmple} during the method's run for the first 120 seconds. $\block{1}$ is the top level block type, $\block{2},\block{3}$ are second level block types, $\block{4}$ is third level block type. Widths of the segments were slightly rescaled to improve readability.}
    \label{fig:dynamics}
\end{figure}

The experiment ran for 10 minutes. The sequence diagram in \cref{fig:dynamics} shows how the decomposition progressed in the first 120 seconds. We can see how the solver first solves the ``master problem'' (first purple bar) for the top block type $\block{1}$, and at 20s it enters its children's subproblems to verify whether the proposed dimensions work or whether cuts need to be introduced. This leads to the same procedure being done in the $\block{2}$, which further calls $\block{4}$ at 27s. Then $\block{3}$ is called at 32s. After that, the ``restricted master problem'' is rapidly solved in $\block{1}$ to obtain the first feasible solution at 33s.

\begin{figure}
    \centering
    \includegraphics[width=0.5\linewidth]{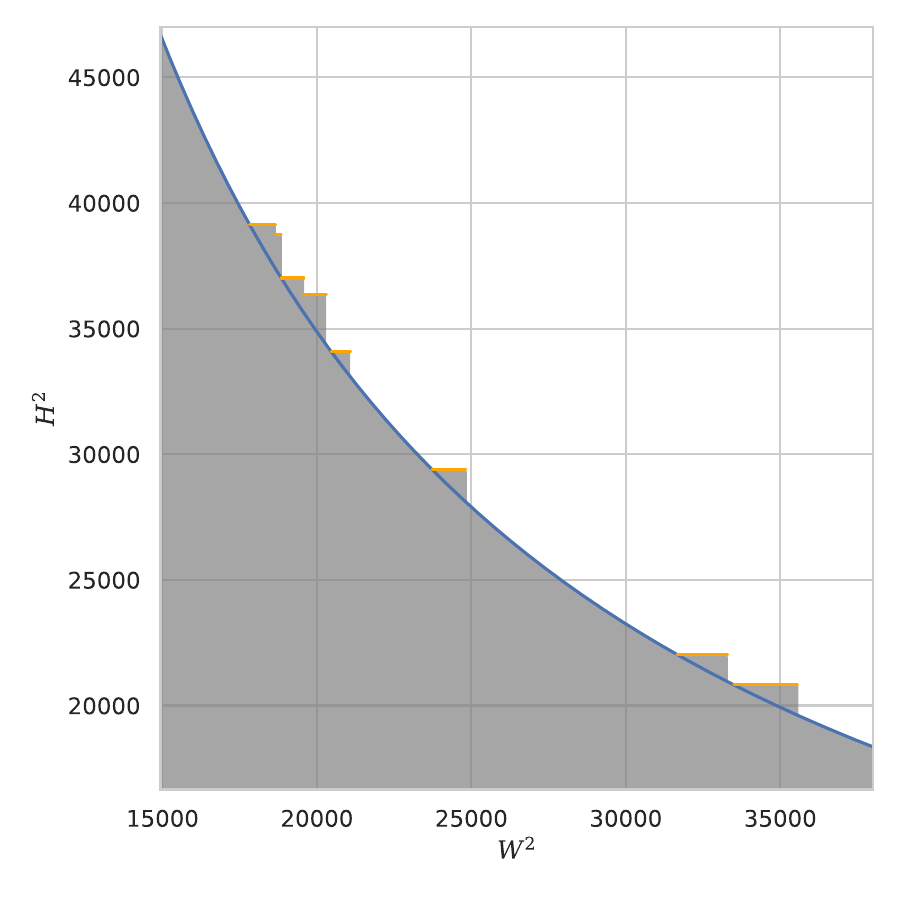}
    \caption{Region of potentially feasible (white) and infeasible (from the perspective of heuristic-LBBD) of $\block{2}$. The original estimate using the lower bound on the area was improved with orange cuts during the experiment.}
    \label{fig:contour-eaxmple}
\end{figure}

In the first 120 seconds, four iterations of the ``master problem'' of $\block{1}$ were finished. In the third iteration from 72s to 100s, the inner loop for the $\block{2}$ was run twice, before the control was returned to $\block{1}$. Furthermore, we can observe how the decomposition searches the possible dimensions of $\block{2}$ in \cref{fig:contour-eaxmple}. We see that $\block{1}$ focused on a solution that utilized tall variants of $\block{2}$, by the number of cuts situated at the left part of \cref{fig:contour-eaxmple}. Two solutions of the 2DHRP problem can be seen in \cref{fig:firstsecond}. The first solution \cref{fig:firstol} was replaced with the latter \cref{fig:secondsol}, since it improved the half-perimeter of the top block type by 490. \cref{fig:secondsol} also shows that the blue block type (including its red child block type) was eventually used in a variant that spans the entire left side of the top block type.

%% file: sections/graph2.tex
\resizebox{0.2\textwidth}{!}{%
\begin{tikzpicture}[
  >={Stealth},
  vertex/.style={circle, draw=black, font=\sffamily\bfseries},
  purple/.style={fill={rgb,255:red,128; green,0; blue,128}, fill opacity=0.65},
  blue/.style={fill={rgb,255:red,30; green,144; blue,255}, fill opacity=0.65},
  green/.style={fill={rgb,255:red,0; green,128; blue,0}, fill opacity=0.65},
  red/.style={fill={rgb,255:red,255; green,0; blue,0}, fill opacity=0.65},
  level distance=15mm
]

\Large
\node[vertex,purple,minimum size=15mm] (A) {$\block{1}$};

\node[vertex,blue,minimum size=15mm,below left=of A] (B) {$\block{2}$};
\node[vertex,green,minimum size=15mm,right=of B] (C) {$\block{3}$};
\node[vertex,red,minimum size=15mm,below=of B] (D) {$\block{4}$};
\graph {
  (A) ->[edge label=1] (B);
  (A) ->[edge label=1, near end] (C);
  (B) ->[edge label=1] (D);
};
\end{tikzpicture}
}

%% file: sections/5-experiments.tex
\section{Experiments}\label{sec:experiments}
We implemented the algorithms using Python 3.10. Experiments were performed on Intel Xeon E5-2690 using a single thread. CP Optimizer v22.1 was used as a CP solver, and Gurobi Optimizer v12.0 as a MILP solver. Implementation is provided in \cite{grus2026hrp}. 

\subsection{Generating Instances}\label{app:gen}
The problem described in this paper does not utilize any standard instance sets found in the literature. Inspired by single-level instances from  \citet{grus-placement}, several sets of instances inspired by the placement of analog integrated circuits were generated to compare the monolithic methods, the existing Bottom-Up approach \citep{xu-placement}, and the proposed heuristic-LBBD method. The generated instance sets are outlined in \cref{tab:instances}, and are provided in \citet{zenodo-grus}. The way the instances were generated is described in the following sections.

\subsubsection{Block Types and Levels}

Each instance set is characterized by the number of levels $l$. Each instance has block types organized in a randomly generated hierarchy. This was done so the maximum path from the top block type to one of its leaves contained exactly $l$ block types (thus, a single-level instance contains only the top block type). We generated instances with up to seven levels, which spans the typical complexity of designed analog integrated circuits.
The average depth of the nodes and the average number of block types per instance are reported in \cref{tab:instances} for each set of generated instances. 

\subsubsection{Rectangles and Block Occurrences}

With the hierarchy determined, rectangles and block occurrences are generated for each node in the graph. One block occurrence per block type was created, given the generated hierarchy. However, for sets \textbf{L3-M} and \textbf{L4-M}, multiple occurrences of the same block type were allowed. The number of rectangles to generate was randomly sampled for each block type. This value was as low as 5 and as high as 40-80 objects (relevant number of components for integrated circuits). 
For instances with more levels (and, thus, implicitly more block types), the upper bound on the number of rectangles was set to the smaller value. The average number of objects per block type and the average total number of objects are shown in \cref{tab:instances}. 

Each rectangle was generated with up to 5 variants (with the exception of \textbf{L1-NV}, where only a single variant was generated) by first sampling an area from the pre-defined interval and then sampling the aspect ratio of the variant. Generation starts at the top block type, and the interval of possible areas is multiplied by a randomly sampled value from the ``area multiplier'' interval $(0.5;1)$ whenever the child block type is recursively entered. This way, the size of the rectangles is reduced for lower-level block types. 

For the two level instances, three sets $\textbf{L2-S},\textbf{L2-I},\textbf{L2-L}$ were generated. The only difference between these sets is the aforementioned "area multiplier" interval. $\textbf{L2-S}$ used interval $(0.1;0.3)$; this made rectangles sampled in the child block type much smaller than those in the parent. $\textbf{L2-I}$ used $(0.3,0.7)$, and $\textbf{L2-L}$ used $(0.7,1.0)$. These instances were used to test whether and how the size of lower-level block types (which depends on the size of their rectangles) affects the optimization.

\subsubsection{Computation Time}
The computation time was fixed for each method given the instance set. The value is reported in the last column of \cref{tab:instances}, and was kept fixed independently of the number of rectangles of the specific instance. For instance, with fewer levels, 10-30 minutes were provided, and a time limit of up to 4 hours was used for the most complex ones.

For multi-level instances, the time is managed as described in \cref{sec:bottom-up} and \cref{sec:proposed-mod}. Time is allocated proportionally among block types for Bottom-Up. LBBD uses an improvement period of 10 seconds before each optimization of the single-block packing problem is aborted (i.e., when the objective does not improve). The main loop for each block type in \cref{fig:diagram} is limited by 30 seconds before the best-so-far solution is ``fine-tuned'' and returned to the parent block type.

 \begin{table}[htbp]
 \centering
 \adjustbox{max width=0.95\textwidth}{%
\begin{tabular}{lcccccccc}\toprule
& instances & avg block types & levels & avg level & avg objects per block type & avg total objects & multiple occurrences & time [min] \\
\midrule
\textbf{L1-NV} & 27 & 1.00 & 1 & 1.00 & 43.78  & 43.78 & No &10\\
\textbf{L1} & 27 & 1.00 & 1 & 1.00 & 43.78  & 43.78 & No & 10\\
\textbf{L2-L} & 21 & 4.57 & 2 & 1.77 & 28.00  & 121.95 & No&10\\
\textbf{L2-I} & 21 & 4.57 & 2 & 1.77 & 28.71  & 124.48 & No &10\\
\textbf{L2-S} & 21 & 4.33 & 2 & 1.75 & 28.59  & 116.67 & No&10\\
\textbf{L3} & 15 & 7.33 & 3 & 2.34 & 19.49  & 136.80 & No&30\\
\textbf{L3-M} & 21 & 7.76 & 3 & 2.27 & 26.35  & 199.76 & Yes&30\\
\textbf{L4} & 15 & 12.40 & 4 & 2.92 & 20.92  & 246.27 & No&120\\
\textbf{L4-M} & 21 & 11.48 & 4 & 2.92 & 26.81  & 288.24 & Yes& 120\\
\textbf{L5} & 21 & 16.76 & 5 & 3.54 & 18.13  & 327.71 & No&120 \\
\textbf{L6} & 21 & 17.76 & 6 & 3.92 & 18.39  & 350.52 & No&240 \\
\textbf{L7} & 18 & 23.44 & 7 & 4.50 & 16.36  & 402.72 & No&240 \\
\end{tabular}}
 \caption{Instance sets and their characteristics.}
 \label{tab:instances}
\end{table}

\subsection{Comparison of the Single-Level Solvers}\label{exp:single-level}
We compared the heuristic baseline $\texttt{HEUR}$ and monolithic models $\texttt{M-CP}$, $\texttt{M-MILP}$ on \textbf{L1-NV} and \textbf{L1} instance sets to determine which of the two solving techniques should be used as the backbone of the decomposition methods. To compare the results, mean (and median in parentheses) values of W+H and AREA gaps across the instance sets are reported in \cref{tab:single-level}. For a given instance and solution with dimensions $W,H$, define:
\begin{equation}
    \mathrm{W+H~gap} = \frac{W + H}{\LBhptop{}} ~[\%]
\end{equation}
\begin{equation}
    \mathrm{AREA~gap} = \frac{W\cdot H}{\LBareatop{}} ~[\%]
\end{equation}

As \cref{tab:single-level} shows, the best results are reported by the \texttt{M-CP} method, both for the area and half-perimeter. The difference is quite significant on $\textbf{L1-NV}$ instances with a single variant per rectangle. The difference is smaller for the multi-variant instances $\textbf{L1}$, but the median of $\texttt{M-CP}$ for the W+H gap is still better: 2.96 instead of 3.97 for $\texttt{M-MILP}$.

An alternative objective for CP was also examined, optimizing the area $W\cdot H$ explicitly. The results in \cref{tab:single-level} for $\texttt{M-CP}_\texttt{AREA}$ show that using the area objective does not help, but rather diminishes the overall performance of the CP solver.

 \begin{table}[htbp]
 \centering
 \adjustbox{max width=0.95\textwidth}{%
\begin{tabular}{lcccccccc}\toprule
& \multicolumn{2}{c}{\texttt{HEUR}} 
& \multicolumn{2}{c}{\texttt{M-MILP}}
& \multicolumn{2}{c}{\texttt{M-CP}}& \multicolumn{2}{c}{$\texttt{M-CP}_\texttt{AREA}$}  \\
\cmidrule(lr){2-3}\cmidrule(lr){4-5}\cmidrule(lr){6-7}\cmidrule(lr){8-9}
    & $\mathrm{W+H~gap}$     & $\mathrm{AREA~gap}$   & $\mathrm{W+H~gap}$     & $\mathrm{AREA~gap}$    & $\mathrm{W+H~gap}$     & $\mathrm{AREA~gap}$  & $\mathrm{W+H~gap}$     & $\mathrm{AREA~gap}$  \\        
\midrule

\textbf{L1-NV} & 9.84 (9.63)& 19.52 (18.68) & 5.99 (6.30) & 10.62 (11.09) & \textbf{4.91 (3.73)} &\textbf{8.76 (6.99)}& 9.52 (7.85) & 12.61 (14.10) \\
\textbf{L1} & 7.17 (6.85) & 14.04 (13.70) & 3.84 (3.97) & 6.72 (6.78) & \textbf{3.63 (2.96)} & \textbf{6.43 (3.76)} & 7.27 (6.56) & 9.21 (10.78)\\
\end{tabular}}
 \caption{Mean (median) percentage values of $\mathrm{W+H~gap}$ and $\mathrm{AREA~gap}$  for different solvers on single-level instances with a time limit of 10 minutes.} 
 \label{tab:single-level}
\end{table}

\begin{figure}
    \centering
    \includegraphics[width=0.5\linewidth]{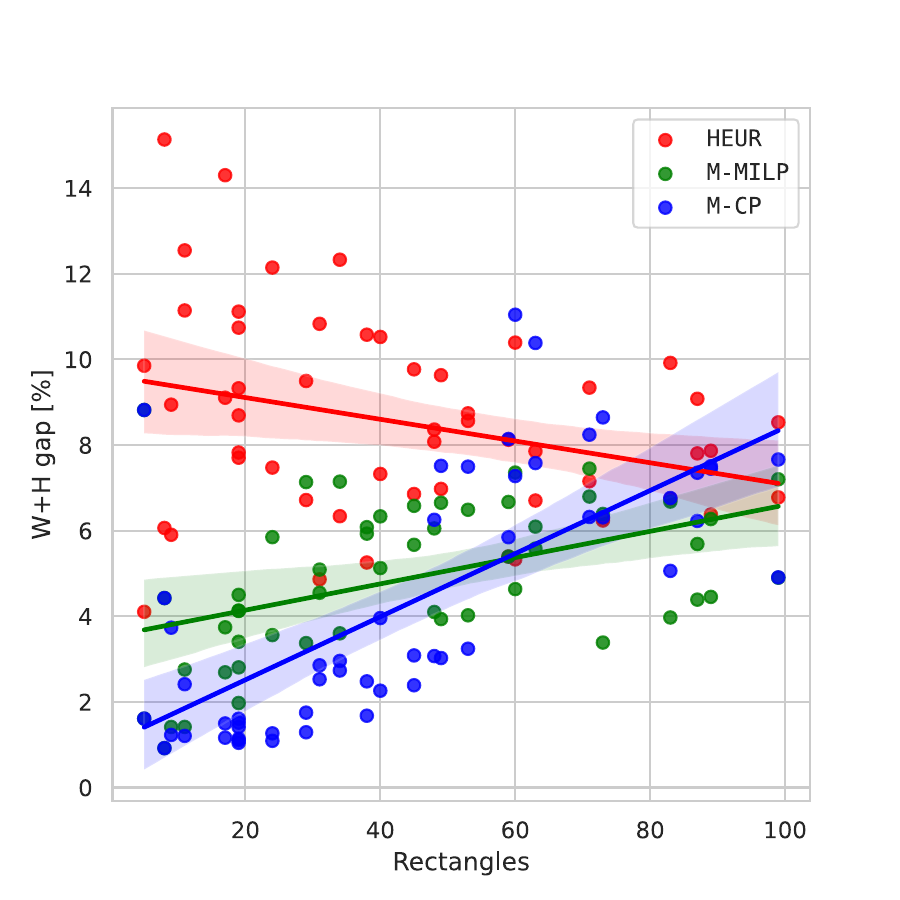}
    \caption{Relationship between the number of rectangles and obtained W+H gap across datasets \textbf{L1-NV} and \textbf{L1} for the three methods, with added regression lines. }
    \label{fig:placeholder}
\end{figure}

The performance according to the size of the instance is visualized in \cref{fig:placeholder}. There, for each instance of $\textbf{L1},\textbf{L1-NV}$, the value of $\mathrm{W+H~gap}$ is shown with respect to the instance's number of rectangles. We can see that the gap reported by the heuristic $\texttt{HEUR}$ improves with increasing number of rectangles, but exact methods are still mostly better, even though their performance worsens. \texttt{M-CP} performs best for the mid-sized instances with 20 - 55 rectangles, but for the larger ones, the \texttt{M-MILP} wins. Since most of the block types of the multiple-level instances were generated with around 50 rectangles, \texttt{M-CP} derived single-block packing solver was used as a solver for both decomposition methods \texttt{BU} and \texttt{LBBD} approaches.

As a final note, the experiment with $\texttt{M-CP}$ and a time limit of 10 hours was performed. The mean values of $\mathrm{W+H~gap}$ were 3.65 for $\textbf{L1-NV}$ (4.91 for 10-minute $\texttt{M-CP}$) and 2.25 for \textbf{L1} (3.63 for 10-minute $\texttt{M-CP}$). Although these values are not obtained from proven optimal solutions, they provide insight into the gap between lower bounds and solutions found by the compared methods.

\subsubsection{Alternative Models}

In our preliminary experiments, we tested several alternative models for CP Optimizer, including omission of the 1D cumulative constraints, explicit pairwise modeling with two $\texttt{noOverlap}$ constraints, and usage of integer variables instead of interval ones. Furthermore, we also experimented with OR-tools CP solver \citep{cpsatlp}, using its $\texttt{noOverlap2D}$ constraint (specific case of \texttt{geost} constraint). However, the model used in this paper provided the overall best results.

\subsection{Multi-Level Instances}\label{sec:multi-level-exp}

In this section, we primarily study the performance of decomposition methods: Bottom-Up \texttt{BU} and LBBD \texttt{LBBD}. \texttt{BU} uses a different number of variants per block type: $\texttt{BU}_3$, $\texttt{BU}_5$, $\texttt{BU}_9$, $\texttt{BU}_{13}$, $\texttt{BU}_{25}$ use 3, 5, 9, 13, and 25 variants, respectively. We also compare them with the monolithic models $\texttt{M-CP}$, $\texttt{M-MILP}$. Heuristic warm starts were provided both for the monolithic models and for individual block types in decomposition methods. 

\texttt{LBBD} versions differ in their use of fine-tuning of \cref{sec:finetune}: $\texttt{LBBD}_0$ does not use the second phase of the fine-tuning at all, $\texttt{BU}_1$ uses the more conservative setting $\alpha=1$, and $\texttt{BU}_\texttt{R}$ uses a more radical version from the end of \cref{sec:finetune}.

Furthermore, values of the ``best bounds'' $\texttt{BU}_\texttt{B}$ and $\texttt{LBBD}_\texttt{B}$ are reported. These are calculated using the best solutions among the respective versions of the method ($\texttt{BU}_3$, $\texttt{BU}_5$, $\texttt{BU}_9$, $\texttt{BU}_{13}$, $\texttt{BU}_{25}$ for $\texttt{BU}_\texttt{B}$; $\texttt{LBBD}_0$, $\texttt{LBBD}_1$, $\texttt{LBBD}_\texttt{R}$ for $\texttt{LBBD}_\texttt{B}$) to calculate the metrics. This is done independently for each instance, and thus $\texttt{BU}_\texttt{B}$ and $\texttt{LBBD}_\texttt{B}$ serve only as best bounds for \texttt{BU} and \texttt{LBBD}.

\subsubsection{Two-Level Instances}

First, we discuss the results on the smallest multi-level instances with two levels only. The results of the directly optimized half-perimeter are reported in \cref{tab:two-level-hp}, and the area values of the same solutions are reported in \cref{tab:two-level-area}.

 \begin{table}[htbp]
 \centering
 \adjustbox{max width=0.99\textwidth}{%
\begin{tabular}{lcc|ccccc|ccc|cc}\toprule
&\texttt{M-MILP} & \texttt{M-CP} & $\texttt{BU}_3$ & $\texttt{BU}_5$ & $\texttt{BU}_9$ & $\texttt{BU}_{13}$ & $\texttt{BU}_{25}$ & $\texttt{LBBD}_0$ & $\texttt{LBBD}_1$ & $\texttt{LBBD}_\texttt{R}$ & $\texttt{BU}_\texttt{B}$ & $\texttt{LBBD}_\texttt{B}$\\
\midrule
\textbf{L2-L} & 6.19 (5.62) & 8.85 (8.62) & 4.45 (4.2) & 4.71 (3.81) & 4.38 (3.82) & 3.49 (3.16) & 3.64 (3.05) & 3.02 (2.85) & 3.04 (2.6) & 3.02 (2.92) & 3.17 (2.92) & \textbf{2.79 (2.45)}\\
\textbf{L2-I} & 6.55 (6.33) & 9.45 (10.29) & 4.29 (4.22) & 4.16 (3.96) & 4.69 (3.87) & 4.0 (3.34) & 4.02 (3.19) & 3.88 (3.42) & 3.55 (3.23) & 3.79 (3.1) & 3.3 (2.78) & \textbf{3.25 (2.93)}\\
\textbf{L2-S} & 6.61 (6.37) & 8.55 (8.99) & 4.11 (3.23) & 4.35 (3.3) & 4.26 (3.51) & 3.21 (2.7) & 3.41 (2.93) & 3.21 (3.05) & 3.2 (2.8) & 3.29 (2.87) & \textbf{2.83 (2.56)} & 2.89 (2.54)\\
\end{tabular}}
 \caption{Mean (median) percentage values of $\mathrm{W+H~gap}$ on two-level instances.}
 \label{tab:two-level-hp}
\end{table}

 \begin{table}[htbp]
 \centering
 \adjustbox{max width=0.99\textwidth}{%
\begin{tabular}{lcc|ccccc|ccc|cc}\toprule
&\texttt{M-MILP} & \texttt{M-CP} & $\texttt{BU}_3$ & $\texttt{BU}_5$ & $\texttt{BU}_9$ & $\texttt{BU}_{13}$ & $\texttt{BU}_{25}$ & $\texttt{LBBD}_0$ & $\texttt{LBBD}_1$ & $\texttt{LBBD}_\texttt{R}$ & $\texttt{BU}_\texttt{B}$ & $\texttt{LBBD}_\texttt{B}$\\
\midrule
\textbf{L2-L} & 11.86 (11.0) & 17.38 (17.64) & 7.96 (7.6) & 8.14 (7.0) & 7.8 (7.66) & 6.39 (6.08) & 6.69 (5.85) & 5.75 (5.45) & 5.7 (5.19) & 5.77 (5.51) & 5.98 (5.6) & \textbf{5.28 (4.65)}\\
\textbf{L2-I} & 12.12 (11.96) & 18.34 (20.43) & 7.96 (7.4) & 7.97 (7.75) & 8.31 (7.35) & 7.46 (5.81) & 7.38 (5.73) & 6.5 (6.14) & 6.47 (6.2) & 6.86 (5.99) & 6.24 (5.49) & \textbf{5.96 (5.78)}\\
\textbf{L2-S} & 12.17 (10.88) & 16.39 (17.74) & 7.41 (5.53) & 8.24 (6.58) & 7.93 (6.25) & 6.0 (5.02) & 6.46 (5.28) & 5.84 (5.39) & 6.01 (4.83) & 6.01 (5.08) & 5.55 (4.94) & \textbf{5.4 (4.58)}\\
\end{tabular}}
 \caption{Mean (median) percentage values  of $\mathrm{AREA~gap}$ on two-level instances.}
 \label{tab:two-level-area}
\end{table}

We can see that the monolithic methods \texttt{M-CP} and \texttt{M-MILP} are not performing well for the multi-level instances. Interestingly, it actually seems that \texttt{M-MILP} outperforms its counterpart, but it is still much worse than decomposition methods. Thus, we do not include them in the comparison later in the paper. Furthermore, differences between the various two-level instance sets are not significant. It suggests that different scaling of the lower-level block type and its rectangles does not have a significant effect on the methods.

If we focus on different versions of \texttt{BU}, it is not straightforward to pinpoint the winner, with the results being very close on all instance sets. The same holds for \texttt{LBBD} versions, and also for the values of half-perimeter and area. However, $\texttt{LBBD}_{\texttt{R}}$ reports non-trivially better results; on average, about 0.3\% smaller $\mathrm{W+H~gap}$ is reported than any $\texttt{BU}$ version; a notable gain given how close both approaches probably are to the theoretical lower bound.


When we compare the ``best bound'' columns $\texttt{BU}_\texttt{B}$ and $\texttt{LBBD}_\texttt{B}$, the differences between the two approaches are not that significant, with $\texttt{BU}_\texttt{B}$ winning for two datasets regarding $\mathrm{W+H~gap}$ by a small margin. This suggests that the Bottom-Up approach is quite powerful, but the incorrect choice of the number of variants to explore negatively affects the individual versions.

\subsubsection{Complex Instances}

\begin{figure}[htbp]
    \centering
    
    \begin{subfigure}[b]{0.99\textwidth}
        \centering
        
    \includegraphics[width=0.5\linewidth]{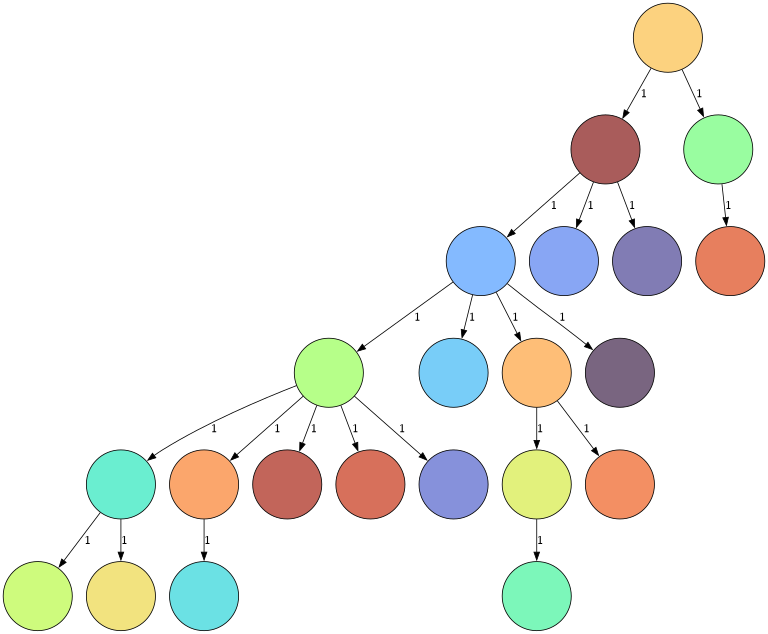}
    \caption{Hierarchy of block types. Nodes corresponding to individual rectangles are omitted.}
    \end{subfigure}
    
    \vskip\baselineskip
    
    
        \begin{subfigure}[b]{0.99\textwidth}  
            \centering 
            
    \includegraphics[width=0.6\linewidth]{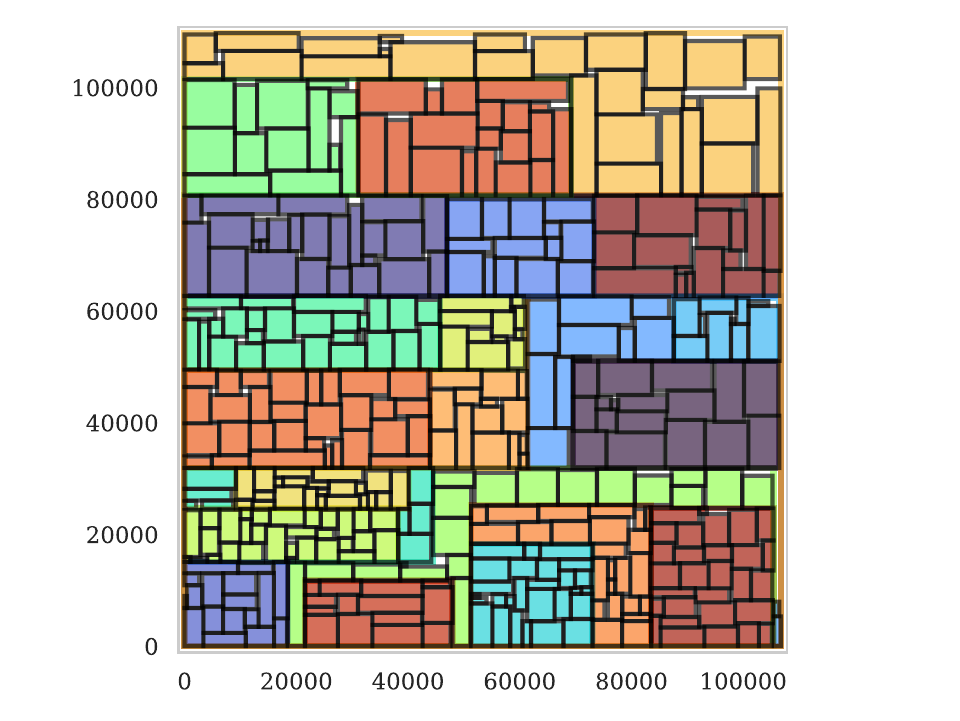}
    \caption{Solution obtained by $\texttt{LBBD}_\texttt{R}$.}
        \end{subfigure}
        \caption{Hierarchy and $\texttt{LBBD}_\texttt{R}$ solution of \textbf{L6} instance. Each block type occurs only once.} 
    \label{fig:sixlevel}
\end{figure}

In \cref{tab:complex-hp,tab:complex-area}, we report the half-perimeter and area results for more complex instances. An example of a complex six-level instance and its solution is shown in \cref{fig:sixlevel}.

 \begin{table}[htbp]
 \centering
 \adjustbox{max width=0.95\textwidth}{%
\begin{tabular}{lccccc|ccc|cc}\toprule
&$\texttt{BU}_3$ & $\texttt{BU}_5$ & $\texttt{BU}_9$ & $\texttt{BU}_{13}$ & $\texttt{BU}_{25}$ & $\texttt{LBBD}_0$ & $\texttt{LBBD}_1$ & $\texttt{LBBD}_\texttt{R}$ & $\texttt{BU}_\texttt{B}$ & $\texttt{LBBD}_\texttt{B}$\\
\midrule
\textbf{L3} & 5.76 (5.18) & 3.84 (3.33) & 3.56 (3.49) & 3.42 (3.34) & 3.53 (2.85) & 2.82 (2.77) & 2.85 (2.89) & 3.03 (2.73) & 2.79 (2.68) & \textbf{2.62 (2.61)}\\
\textbf{L3-M} & 8.66 (6.53) & 4.88 (4.93) & 5.34 (4.72) & 4.35 (4.04) & 4.57 (3.64) & 4.2 (3.76) & 3.99 (3.7) & 4.11 (3.79) & 3.7 (3.42) & \textbf{3.66 (3.37)}\\
\textbf{L4} & 7.04 (6.44) & 4.18 (3.84) & 3.67 (3.34) & 3.65 (3.52) & 3.23 (3.2) & 3.29 (3.22) & 3.32 (2.85) & 3.13 (3.11) & \textbf{2.83 (2.85)} & 2.96 (2.85)\\
\textbf{L4-M} & 16.32 (14.8) & 8.57 (7.32) & 7.11 (6.57) & 8.24 (5.7) & 6.61 (4.15) & 5.21 (3.98) & 5.18 (4.03) & 5.32 (4.12) & 5.16 (4.05) & \textbf{4.54 (3.64)}\\
\textbf{L5} & 12.73 (13.35) & 5.2 (5.05) & 4.39 (3.93) & 4.31 (4.3) & 4.56 (3.97) & 3.32 (3.17) & 3.46 (3.25) & 3.61 (3.45) & 3.65 (3.51) & \textbf{3.2 (3.17)}\\
\textbf{L6} & 17.66 (16.52) & 6.3 (6.71) & 4.27 (4.19) & 4.16 (4.16) & 4.14 (4.09) & 3.39 (3.32) & 3.45 (3.35) & 3.55 (3.42) & 3.59 (3.35) & \textbf{3.22 (3.17)}\\
\textbf{L7} & 53.58 (31.09) & 10.51 (7.7) & 7.93 (4.91) & 7.13 (4.62) & 6.72 (4.66) & 3.8 (3.56) & 3.92 (3.76) & 3.83 (3.58) & 6.23 (4.24) & \textbf{3.59 (3.4)}\\
\end{tabular}}
 \caption{Mean (median) percentage values of $\mathrm{W+H~gap}$ on complex multi-level instances.}
 \label{tab:complex-hp}
\end{table}

 \begin{table}[htbp]
 \centering
 \adjustbox{max width=0.95\textwidth}{%
\begin{tabular}{lccccc|ccc|cc}\toprule
&$\texttt{BU}_3$ & $\texttt{BU}_5$ & $\texttt{BU}_9$ & $\texttt{BU}_{13}$ & $\texttt{BU}_{25}$ & $\texttt{LBBD}_0$ & $\texttt{LBBD}_1$ & $\texttt{LBBD}_\texttt{R}$ & $\texttt{BU}_\texttt{B}$ & $\texttt{LBBD}_\texttt{B}$\\
\midrule
\textbf{L3} & 10.81 (9.27) & 7.11 (6.67) & 6.65 (6.87) & 6.33 (6.57) & 6.1 (5.48) & 5.36 (5.22) & 5.41 (5.31) & 5.64 (5.33) & 5.48 (5.27) & \textbf{5.12 (5.19)}\\
\textbf{L3-M} & 16.09 (12.76) & 9.37 (8.52) & 9.49 (8.72) & 8.09 (6.91) & 8.46 (6.98) & 7.86 (6.9) & 7.52 (7.05) & 7.98 (6.81) & \textbf{7.05 (6.36)} & 7.17 (6.53)\\
\textbf{L4} & 13.58 (13.08) & 7.98 (7.6) & 6.92 (6.53) & 6.53 (6.69) & 5.71 (5.6) & 6.36 (6.09) & 6.55 (5.62) & 5.96 (5.69) & \textbf{5.32 (5.32)} & 5.61 (5.09)\\
\textbf{L4-M} & 30.85 (28.97) & 14.96 (12.86) & 12.13 (11.66) & 13.62 (11.38) & 11.09 (8.22) & 9.25 (7.63) & 9.14 (7.32) & 9.16 (7.64) & 9.38 (7.77) & \textbf{8.45 (6.98)}\\
\textbf{L5} & 25.4 (25.79) & 10.14 (9.68) & 8.4 (8.02) & 8.34 (7.74) & 8.11 (7.41) & 6.53 (6.37) & 6.67 (6.3) & 6.92 (6.33) & 7.26 (7.1) & \textbf{6.27 (6.05)}\\
\textbf{L6} & 34.27 (30.56) & 12.18 (12.59) & 8.35 (7.58) & 7.98 (8.23) & 8.03 (7.52) & 6.61 (6.42) & 6.79 (6.32) & 6.98 (6.71) & 7.26 (6.76) & \textbf{6.36 (6.27)}\\
\textbf{L7} & 80.59 (62.32) & 16.28 (13.87) & 12.11 (9.61) & 10.62 (9.44) & 9.77 (9.4) & 7.21 (7.12) & 7.44 (7.57) & 7.54 (7.22) & 9.21 (8.51) & \textbf{6.94 (6.71)}\\
\end{tabular}}
 \caption{Mean (median) percentage values of $\mathrm{AREA~gap}$ on complex multi-level instances.}
 \label{tab:complex-area}
\end{table}

As before, \cref{tab:complex-hp} shows how the ``best bound'' $\texttt{LBBD}_\texttt{B}$ outperforms its Bottom-Up counterpart $\texttt{BU}_\texttt{B}$. The difference between their  $\mathrm{W+H~gap}$ is between -0.1 percentage point and 2.5 percentage points for $\textbf{L7}$.

\begin{figure}[htbp]
    \centering
    
    \begin{subfigure}[b]{0.5\textwidth}
        \centering
        \includegraphics[width=0.6\textwidth]{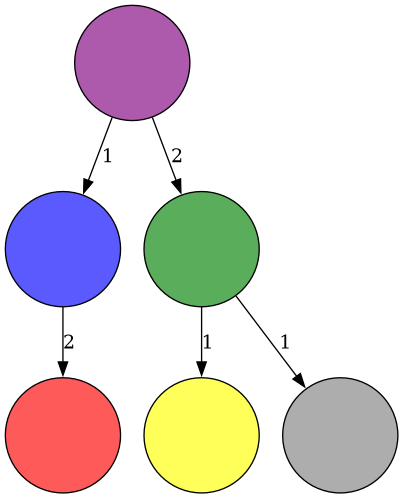}
        \caption{Hierarchy of block types. Nodes corresponding to individual rectangles are omitted. Red and green block types are used multiple times.}
        \label{fig:top}
    \end{subfigure}
    
    \vskip\baselineskip
    
    
        \begin{subfigure}[b]{0.49\textwidth}  
            \centering 
            \includegraphics[width=\textwidth]{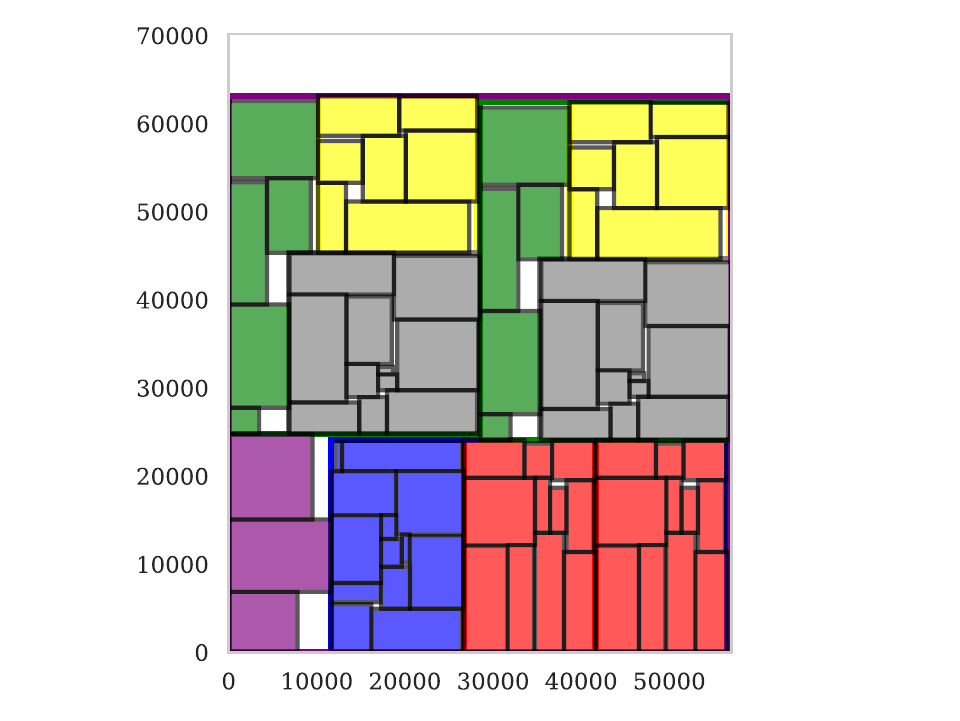}
            \caption{$\texttt{BU}_{13}$: W+H gap = 4.03\%, AREA gap = 7.96\%} 
            \label{fig:bu}
        \end{subfigure}
        \hfill
        \begin{subfigure}[b]{0.49\textwidth}
            \centering
            \includegraphics[width=\textwidth]{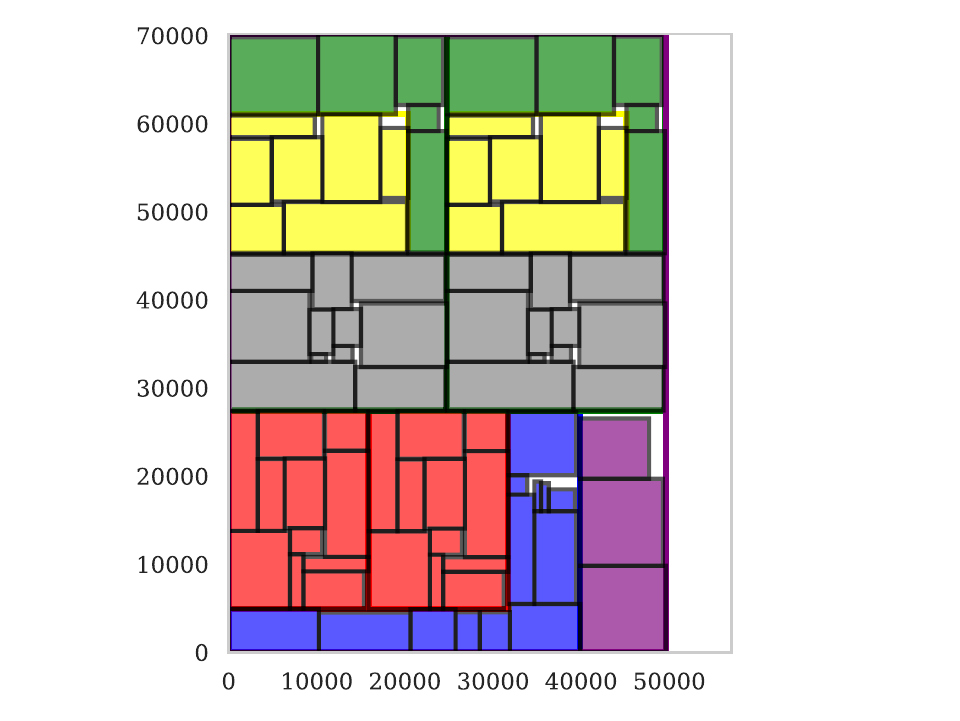}
            \caption{$\texttt{LBBD}_{\texttt{R}}$: W+H gap = 3.68\%, AREA gap = 4.32\%}    
            \label{fig:LBBD}
        \end{subfigure}
        \caption{Comparison of solutions obtained for instance of \textbf{L3-M}. Notice that the red block type and the green block type (which contains grey and yellow) are used multiple times.} 
        \label{fig:buLBBD}
\end{figure}

The difference seems to be larger for \textbf{L3-M} and \textbf{L4-M}, where multiple occurrences of the same block type are used. This suggests that the multiple occurrences of the ``nonoptimally'' packed block type may lead to multiplication of the wasted space in the upper levels. Visually, this is presented for one instance of \textbf{L3-M} in \cref{fig:buLBBD}, where solutions obtained by the best performing version of each method are shown. Thus, a more informed approach that can reason about a block type's dimensions beforehand may be much better suited for such a scenario.

For set \textbf{L4}, the ``best bound'' $\texttt{BU}_\texttt{B}$ outperforms the ``best bound'' for LBBD. However, when considering individual LBBD and Bottom-Up variants, $\texttt{LBBD}_\texttt{R}$ provides the best performance. The stronger aggregated result of $\texttt{BU}_\texttt{B}$ stems from the complementarity of different Bottom-Up configurations on this dataset. In particular, $\texttt{BU}_3$ and $\texttt{BU}_5$ methods, which consider fewer variants per rectangle, produce solutions with $\mathrm{W+H~gap}$ that is a few percentage points lower than that of $\texttt{BU}_{25}$ on several instances. Together, they yield better aggregated results for this instance set. For LBBD, such complementarity is not observed, as the performance of different configurations varies only marginally across instances.

If we focus only on individual versions of both methods, we can see that the proposed method performs better, and the difference between them seems to increase with the number of levels, up to $2.5$ percentage points for set \textbf{L7}. It can be clearly seen that the small number of solutions generated by $\texttt{BU}_3$ is detrimental, since there is no guarantee that a suitable partial packing was produced throughout the hierarchy. However, not even $\texttt{BU}_{13}$ or $\texttt{BU}_{25}$ are good enough to defeat \texttt{LBBD} methods, probably due to the amount of time wasted on optimizing useless variants. 

\begin{figure}
    \centering
    \includegraphics[width=0.7\linewidth]{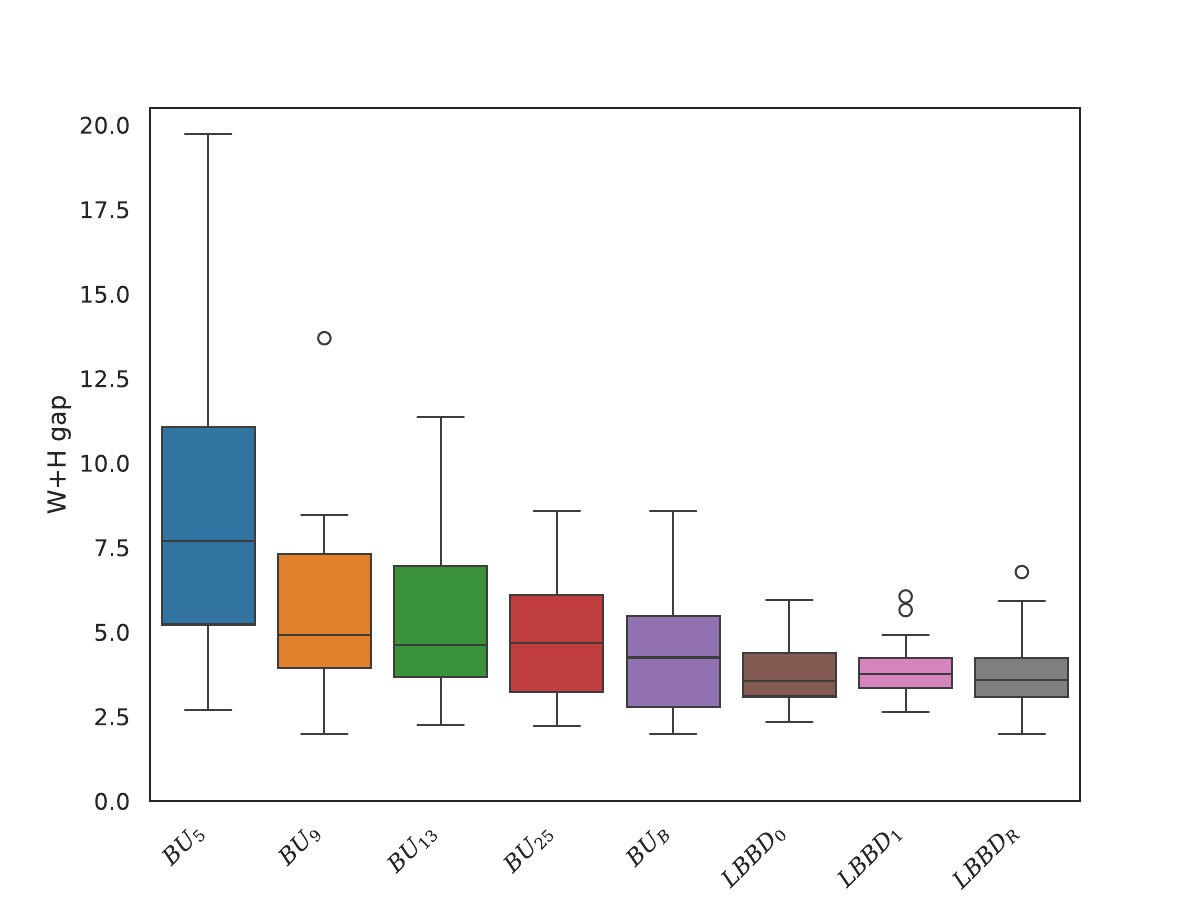}
    \caption{Box plot of W+H gap for instances of \textbf{L7}.}
    \label{fig:placeholder5}
\end{figure}

This can be further studied in the box plot generated for instances of \textbf{L7}, shown in \cref{fig:placeholder5}. It is clear that the width of the inter-quartile range is much greater for the versions of \texttt{BU} in comparison to versions of \texttt{LBBD}. This suggests that \texttt{LBBD} produces more consistent results. When the different versions of \texttt{LBBD} are compared, there is not much difference between them with respect to the results. Thus, it may not be necessary to strengthen the cuts generated by subproblems, at least for the instances considered in this paper, simplifying the overall method.

    Altogether, the results suggest that the proposed LBBD method performs better on the generated instances. This is especially true for more complex instances with more block types spread across more levels, as the results for sets $\textbf{L6},\textbf{L7}$ demonstrated. However, we need to consider that both LBBD and Bottom-up methods are finding a solution very close to the theoretical lower bound, and therefore, even the straightforward Bottom-Up method is a suitable way to tackle the 2DHRP. 

\subsubsection{Parameter Sensitivity}
From the results presented in the \cref{sec:multi-level-exp}, recommendations on several key hyperparameters of the used methods can be derived. With respect to the proposed time management strategy, the Bottom-Up method highly benefits from the provided number of variants $N$. Both $N=13$ and $N=25$ provided the overall best results across all instance sets, with $\texttt{BU}_{25}$ working the best for the instances with most levels and block types. This suggests that the large variety of solutions outperforms the smaller selection of well-optimized variants ($\texttt{BU}_{25}$ essentially has half the time to optimize the given variant in comparison with  $\texttt{BU}_{13}$). It is therefore crucial to use as many variants as possible, given that the initial optimization of the variant can still be done in a limited time (i.e., the first few moments when the solution's objective is rapidly decreased). 

In the context of LBBD, we consider the effect of a variant of the method with respect to strengthening the cuts ``to the right'', as was described in \cref{sec:finetune}. We decided among $\texttt{LBBD}_0$, which does not perform the strengthening, $\texttt{LBBD}_1$, which tries to find a solution with height one less than the current one, and $\texttt{LBBD}_\texttt{R}$ that uses the height decrement $\alpha$. In addition to $\alpha=0.05 \cdot H^i_\texttt{ACT}$, we consider  $\alpha=0.2 \cdot H^i_\texttt{ACT}$ and $\alpha=0.4 \cdot H^i_\texttt{ACT}$ as well. Note that the cut reduces the search space more with a larger height decrement, but it eliminates otherwise feasible solutions from the search space.

Interestingly, for almost all of the discussed datasets, more extreme steps $0.4$ and $0.2$ performed slightly better; the actual differences between averaged and median values of $\mathrm{W+H~gap}$ of these methods were 0.5$\%$ at most. When aggregated across all instance sets, the best performing variant with $\alpha=0.2 \cdot H^i_\texttt{ACT}$ achieved a mean rank - when methods were sorted by their objective on given instance - of 2.58 (the worst method had 3.16) and a mean $\mathrm{W+H~gap}$ was 3.39 (the worst was 3.70). Therefore, the selection of the LBBD variant does not really influence the result; it seems that the more heuristic the cut becomes, the slightly better results are achieved (with respect to the limited computation time).

\subsection{Exact LBBD}

In this section, the performance of the exact LBBD is briefly examined. It follows the outlined method in \cref{sec:summary} with the single-block packing problem being always run until optimality. Since the original method's cuts are very conservative, we also utilize strengthened cuts from \cref{sec:finetune} that do not affect optimality. Finally, we calculate the upper bound at each iteration by replacing the block types' approximation with the solution provided by the child subproblem for visualization purposes.

For the method to run exactly and with reasonable computation time, it is necessary to prove optimality in each single-block packing problem in a short amount of time. From experiments in \cref{exp:single-level}, it turned out that both CP and MILP solvers struggle to close the gap with just around 12 rectangles per block type within 10 minutes. Therefore, we limit the experiments to a newly generated set of 2,3, and 4- level instances with at most 10 objects per block type in this section (15 in total).

\begin{figure}
    \centering
\includegraphics[width=0.7\linewidth]{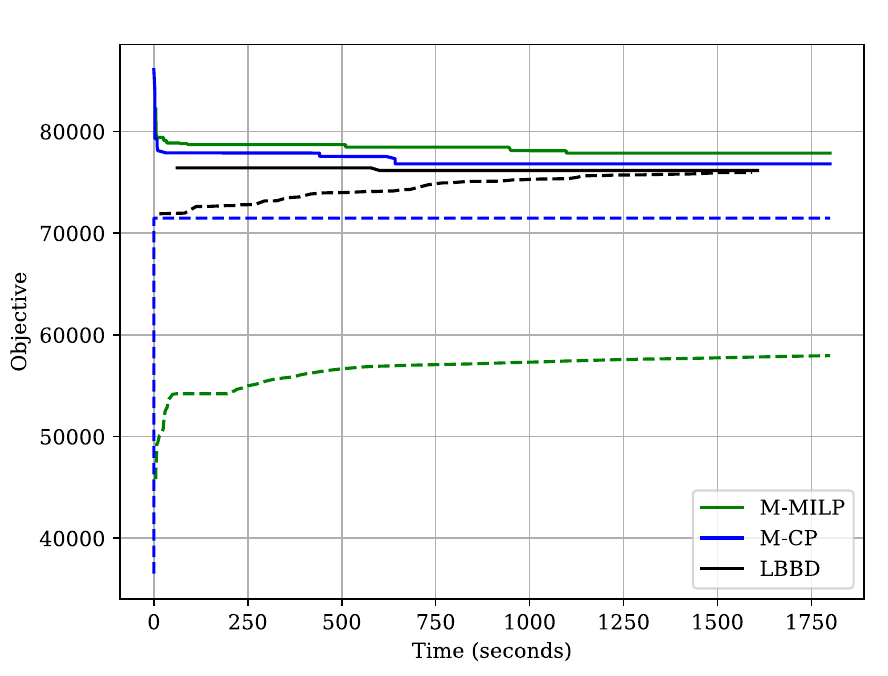}
    \caption{Convergence of UB (solid) and LB (dashed) values of different solvers for a small-scale three-level instance.}
    \label{fig:theoretical}
\end{figure}

For such instances, \texttt{LBBD} is able to reduce the gap between the lower bound and the upper bound. This is shown for a single instance in \cref{fig:theoretical} where it outperforms both $\texttt{M-MILP}$ and $\texttt{M-CP}$. Altogether, the mean (median) percentage gap after 30 minutes of computation was 1.03 (0.00) for \texttt{LBBD}, 16.15 (14.08) for \texttt{M-MILP}, and 3.01 (0.01) for \texttt{M-CP}. This suggests that the exact LBBD is better on these special, smaller-scale instances than monolithic models. However, for a 4-level instance with up to 10 rectangles per block type, \texttt{LBBD} could not finish even a single iteration of the main loop for the top block type (being stuck iterating in lower levels), which shows the necessity of the heuristic modifications suggested in this paper.

%% file: sections/6-discussion-conclusion.tex
\section{Discussion}\label{sec:discussion}
The proposed method seems to, on average, outperform the Bottom-Up method for a diverse set of instances. Furthermore, it is solver independent, since any modeling method and appropriate solver could be utilized to solve the single-block packing problem, as long as the width-and-height cuts can be generated and added to the solver. This could be extended to 3D to apply a similar decomposition method for more practical logistics problems.

Proposed decomposition could potentially be utilized on single-level packing instances. If the complexity of such an instance is too high, a ``virtual hierarchy'' can be artificially created by clustering some rectangles into ``virtual'' block types. While this may sacrifice optimality, since the optimal solution of the original (non-clustered) problem would have dimensions less than or equal to those of the clustered version, the computational performance provided by the hierarchical framework could be improved. Conversely, by removing the explicit boundary of the block types of the hierarchy, a more efficient solution could be obtained. However, this approach eliminates structural constraints such as the compactness of block types. In application domains such as integrated circuit design, enforcing well-defined rectangular interfaces is often essential for modularity and reusability.

For practical application, such as in the design of integrated circuits of \citet{xu-placement,grus-placement}, some other objectives need to be considered (e.g., length of the components interconnections), but also complex constraints regarding, for example, non-uniform minimum distances between rectangles. Both of these problems complicate the way the cuts are generated, and how to evaluate the quality of the child subproblem solutions, and pass such information back to the parent.

\section{Conclusion}\label{sec:conclusion}

In this paper, we formalized a hierarchical packing problem, which models the core features of the packing and placements problems found in (i) design of integrated circuits, (ii) design and planning of facility layouts, and (iii) packing in logistics. 

Due to the complexity of the problem, we implemented a baseline Bottom-Up method, and we proposed a recursive LBBD method. The main advantage of the proposed method is that it is up to the parent block type to select a suitable dimension of the subblock types, rather than randomly generating them as in the Bottom-Up method.

We evaluated the MILP and CP solvers on single-level packing instances and used the CP solver further within the decomposition methods due to its performance. When we compared the Bottom-Up and LBBD methods on instances with between two and seven levels, we showed that our proposed LBBD is superior to the Bottom-Up.

%% file: biblio.bib
@inproceedings{data-drive-multi-elvel,
author = {Chen, Lei and Tong, Xialiang and Yuan, Mingxuan and Zeng, Jia and Chen, Lei},
title = {A Data-Driven Approach for Multi-level Packing Problems in Manufacturing Industry},
year = {2019},
isbn = {9781450362016},
publisher = {Association for Computing Machinery},
address = {New York, NY, USA},
url = {https://doi.org/10.1145/3292500.3330708},
doi = {10.1145/3292500.3330708},
booktitle = {Proceedings of the 25th ACM SIGKDD International Conference on Knowledge Discovery \& Data Mining},
pages = {1762–1770},
numpages = {9},
location = {Anchorage, AK, USA},
series = {KDD '19}
}

@article{ZHANG2025,
title = {An Exact Algorithm for Placement Optimization in Circuit Design},
journal = {Engineering},
year = {2025},
issn = {2095-8099},
doi = {https://doi.org/10.1016/j.eng.2025.03.020},
url = {https://www.sciencedirect.com/science/article/pii/S2095809925001687},
author = {Binqi Zhang and Lu Zhen and Gilbert Laporte}
}

@ARTICLE{zhu-circuits,
  author={Zhu, Keren and Chen, Hao and Liu, Mingjie and Pan, David Z.},
  journal={IEEE Transactions on Computer-Aided Design of Integrated Circuits and Systems}, 
  title={Hierarchical Analog and Mixed-Signal Circuit Placement Considering System Signal Flow}, 
  year={2023},
  volume={42},
  number={8},
  pages={2689-2702},
  doi={10.1109/TCAD.2022.3230367}}

@article{cintra-columns,
title = {Algorithms for two-dimensional cutting stock and strip packing problems using dynamic programming and column generation},
journal = {European Journal of Operational Research},
volume = {191},
number = {1},
pages = {61-85},
year = {2008},
issn = {0377-2217},
doi = {https://doi.org/10.1016/j.ejor.2007.08.007},
url = {https://www.sciencedirect.com/science/article/pii/S0377221707008831},
author = {G.F. Cintra and F.K. Miyazawa and Y. Wakabayashi and E.C. Xavier}
}

@article{pisinger-binpacking,
author = {Pisinger, David and Sigurd, Mikkel},
year = {2007},
month = {02},
pages = {36-51},
title = {Using Decomposition Techniques and Constraint Programming for Solving the Two-Dimensional Bin-Packing Problem},
volume = {19},
journal = {INFORMS Journal on Computing},
doi = {10.1287/ijoc.1060.0181}
}

@article{facility-exact,
title = {Optimal plant layout considering the safety instrumented system design for hazardous equipment},
journal = {Process Safety and Environmental Protection},
volume = {124},
pages = {97-120},
year = {2019},
issn = {0957-5820},
doi = {https://doi.org/10.1016/j.psep.2019.01.021},
url = {https://www.sciencedirect.com/science/article/pii/S095758201831005X},
author = {Julio A. {de Lira-Flores} and Antioco López-Molina and Claudia Gutiérrez-Antonio and Richart Vázquez-Román}
}

@article{cote-cuts,
 ISSN = {0030364X, 15265463},
 URL = {http://www.jstor.org/stable/24540603},
 author = {Jean-François Côté and Mauro Dell'Amico and Manuel Iori},
 journal = {Operations Research},
 number = {3},
 pages = {643--661},
 publisher = {INFORMS},
 title = {Combinatorial Benders' Cuts for the Strip Packing Problem},
 urldate = {2025-06-16},
 volume = {62},
 year = {2014}
}

@article{cote-cuts2,
  title={A Primal Decomposition Algorithm for the Two-dimensional Bin Packing Problem},
  author={Jean-François C{\^o}t{\'e} and Mohamed Haouari and Manuel Iori},
  journal={ArXiv},
  year={2019},
  volume={abs/1909.06835},
  url={https://api.semanticscholar.org/CorpusID:202577569}
}

@misc{Blansch_2022_bachelor, title={Solving the multi-level bin packing problem with time windows using Integer Programming}, url={https://resolver.tudelft.nl/cace05a9-aa09-4226-bf93-099308331ac2}, journal={TU Delft}, author={Le Blansch, Max}, year={2022}}

@article{gomory-2d,
 ISSN = {0030364X, 15265463},
 URL = {http://www.jstor.org/stable/167956},
 author = {P. C. Gilmore and R. E. Gomory},
 journal = {Operations Research},
 number = {1},
 pages = {94--120},
 publisher = {INFORMS},
 title = {Multistage Cutting Stock Problems of Two and More Dimensions},
 urldate = {2025-07-02},
 volume = {13},
 year = {1965}
}

@article{gomory-1d,
  title = {A Linear Programming Approach to the Cutting-Stock Problem},
  volume = {9},
  ISSN = {1526-5463},
  url = {http://dx.doi.org/10.1287/opre.9.6.849},
  DOI = {10.1287/opre.9.6.849},
  number = {6},
  journal = {Operations Research},
  publisher = {Institute for Operations Research and the Management Sciences (INFORMS)},
  author = {Gilmore,  P. C. and Gomory,  R. E.},
  year = {1961},
  month = dec,
  pages = {849–859}
}

@article{heur-survey,
  title={A SURVEY ON HEURISTICS FOR THE TWO-DIMENSIONAL RECTANGULAR STRIP PACKING PROBLEM},
  author={Jose Fernando Oliveira and Alvaro Neuenfeldt J{\'u}nior and Elsa Silva and Maria Ant{\'o}nia Carravilla},
  journal={Pesquisa Operacional},
  year={2016},
  volume={36},
  pages={197-226},
  url={https://api.semanticscholar.org/CorpusID:126170489}
}

@article{altogether-survey,
author = {Oliveira, Oscar and Gamboa, Dorabela and Silva, Elsa},
title = {An introduction to the two-dimensional rectangular cutting and packing problem},
journal = {International Transactions in Operational Research},
volume = {30},
number = {6},
pages = {3238-3266},
doi = {https://doi.org/10.1111/itor.13236},
url = {https://onlinelibrary.wiley.com/doi/abs/10.1111/itor.13236},
eprint = {https://onlinelibrary.wiley.com/doi/pdf/10.1111/itor.13236},
year = {2023}
}

@article{exact-survey,
title = {Exact solution techniques for two-dimensional cutting and packing},
journal = {European Journal of Operational Research},
volume = {289},
number = {2},
pages = {399-415},
year = {2021},
issn = {0377-2217},
doi = {https://doi.org/10.1016/j.ejor.2020.06.050},
url = {https://www.sciencedirect.com/science/article/pii/S0377221720306111},
author = {Manuel Iori and Vinícius L. {de Lima} and Silvano Martello and Flávio K. Miyazawa and Michele Monaci}
}

@article{space-indexed-marecek,
title = {A space-indexed formulation of packing boxes into a larger box},
journal = {Operations Research Letters},
volume = {40},
number = {1},
pages = {20-24},
year = {2012},
issn = {0167-6377},
doi = {https://doi.org/10.1016/j.orl.2011.10.008},
url = {https://www.sciencedirect.com/science/article/pii/S0167637711001131},
author = {Sam D. Allen and Edmund K. Burke and Jakub Mareček}
}

@article{space-index-beasley,
 ISSN = {0030364X, 15265463},
 URL = {http://www.jstor.org/stable/170866},
 author = {J. E. Beasley},
 journal = {Operations Research},
 number = {1},
 pages = {49--64},
 publisher = {INFORMS},
 title = {An Exact Two-Dimensional Non-Guillotine Cutting Tree Search Procedure},
 urldate = {2025-07-02},
 volume = {33},
 year = {1985}
}

@InProceedings{pre-korf,
author="Berger, Martin
and Schr{\"o}der, Michael
and K{\"u}fer, Karl-Heinz",
editor="Fleischmann, Bernhard
and Borgwardt, Karl-Heinz
and Klein, Robert
and Tuma, Axel",
title="A Constraint-Based Approach for the Two-Dimensional Rectangular Packing Problem with Orthogonal Orientations",
booktitle="Operations Research Proceedings 2008",
year="2009",
publisher="Springer Berlin Heidelberg",
address="Berlin, Heidelberg",
pages="427--432",
isbn="978-3-642-00142-0"
}

@article{relative-container,
title = {An analytical model for the container loading problem},
journal = {European Journal of Operational Research},
volume = {80},
number = {1},
pages = {68-76},
year = {1995},
issn = {0377-2217},
doi = {https://doi.org/10.1016/0377-2217(94)00002-T},
url = {https://www.sciencedirect.com/science/article/pii/037722179400002T},
author = {C.S. Chen and S.M. Lee and Q.S. Shen}
}

@Article{Korf2010,
author={Korf, Richard E.
and Moffitt, Michael D.
and Pollack, Martha E.},
title={Optimal rectangle packing},
journal={Annals of Operations Research},
year={2010},
month={Sep},
day={01},
volume={179},
number={1},
pages={261-295},
issn={1572-9338},
doi={10.1007/s10479-008-0463-6},
url={https://doi.org/10.1007/s10479-008-0463-6}
}

@article{logic-based-graph,
title = {Logic based Benders' decomposition for orthogonal stock cutting problems},
journal = {Computers \& Operations Research},
volume = {78},
pages = {290-298},
year = {2017},
issn = {0305-0548},
doi = {https://doi.org/10.1016/j.cor.2016.09.009},
url = {https://www.sciencedirect.com/science/article/pii/S0305054816302301},
author = {Maxence Delorme and Manuel Iori and Silvano Martello}
}

@article{temp-1d-binpack,
title = {A branch-and-price algorithm for the temporal bin packing problem},
journal = {Computers \& Operations Research},
volume = {114},
pages = {104825},
year = {2020},
issn = {0305-0548},
doi = {https://doi.org/10.1016/j.cor.2019.104825},
url = {https://www.sciencedirect.com/science/article/pii/S0305054819302679},
author = {Mauro Dell’Amico and Fabio Furini and Manuel Iori}
}

@Article{alvarez-lb,
author={Alvarez-Valdes, R.
and Parre{\~{n}}o, F.
and Tamarit, J. M.},
title={A branch and bound algorithm for the strip packing problem},
journal={OR Spectrum},
year={2009},
month={Apr},
day={01},
volume={31},
number={2},
pages={431-459},
issn={1436-6304},
doi={10.1007/s00291-008-0128-5},
url={https://doi.org/10.1007/s00291-008-0128-5}
}

@article{martello-lb,
author = {Martello, Silvano and Monaci, Michele and Vigo, Daniele},
year = {2003},
month = {08},
pages = {310-319},
title = {An Exact Approach to the Strip-Packing Problem},
volume = {15},
journal = {INFORMS Journal on Computing},
doi = {10.1287/ijoc.15.3.310.16082}
}

@article{Boschetti-lb,
  title = {An Exact Algorithm for the Two-Dimensional Strip-Packing Problem},
  volume = {58},
  ISSN = {1526-5463},
  url = {http://dx.doi.org/10.1287/opre.1100.0833},
  DOI = {10.1287/opre.1100.0833},
  number = {6},
  journal = {Operations Research},
  publisher = {Institute for Operations Research and the Management Sciences (INFORMS)},
  author = {Boschetti,  Marco Antonio and Montaletti,  Lorenza},
  year = {2010},
  month = dec,
  pages = {1774–1791}
}

@article{martello-minimum-square,
title = {Models and algorithms for packing rectangles into the smallest square},
journal = {Computers \& Operations Research},
volume = {63},
pages = {161-171},
year = {2015},
issn = {0305-0548},
doi = {https://doi.org/10.1016/j.cor.2015.04.024},
url = {https://www.sciencedirect.com/science/article/pii/S0305054815001161},
author = {Silvano Martello and Michele Monaci}
}

@inproceedings{xu-placement,
author = {Xu, Biying and Li, Shaolan and Xu, Xiaoqing and Sun, Nan and Pan, David Z.},
title = {Hierarchical and Analytical Placement Techniques for High-Performance Analog Circuits},
year = {2017},
isbn = {9781450346962},
publisher = {Association for Computing Machinery},
address = {New York, NY, USA},
url = {https://doi.org/10.1145/3036669.3036678},
doi = {10.1145/3036669.3036678},
booktitle = {Proceedings of the 2017 ACM on International Symposium on Physical Design},
pages = {55–62},
numpages = {8},
location = {Portland, Oregon, USA},
series = {ISPD '17}
}

@Article{kubalik,
AUTHOR = {Kubalík, Jiří and Kurilla, Lukáš and Kadera, Petr},
TITLE = {Facility Layout Problem with Alternative Facility Variants},
JOURNAL = {Applied Sciences},
VOLUME = {13},
YEAR = {2023},
NUMBER = {8},
ARTICLE-NUMBER = {5032},
URL = {https://www.mdpi.com/2076-3417/13/8/5032},
ISSN = {2076-3417},
DOI = {10.3390/app13085032}
}

@article{lewis-trapezoid,
title = {Exact algorithms in bar nesting: How to cut general items from linear stocks so that wastage is minimised},
journal = {Computers \& Industrial Engineering},
volume = {200},
pages = {110838},
year = {2025},
issn = {0360-8352},
doi = {https://doi.org/10.1016/j.cie.2024.110838},
url = {https://www.sciencedirect.com/science/article/pii/S0360835224009604},
author = {Rhyd Lewis and Louis Bonnet}
}

@article{grus-perioic,
title = {Periodic chains scheduling on dedicated resources - A crucial problem in time-sensitive networks},
journal = {Computers \& Operations Research},
volume = {180},
pages = {107072},
year = {2025},
issn = {0305-0548},
doi = {https://doi.org/10.1016/j.cor.2025.107072},
url = {https://www.sciencedirect.com/science/article/pii/S0305054825001005},
author = {Josef Grus and Claire Hanen and Zdeněk Hanzálek}
}

@article{bottomleftfill,
  title={The Bottomn-Left Bin-Packing Heuristic: An Efficient Implementation},
  author={Bernard Chazelle},
  journal={IEEE Transactions on Computers},
  year={1983},
  volume={C-32},
  pages={697-707},
  url={https://api.semanticscholar.org/CorpusID:7348178}
}

@article{bestfit,
title = {The best-fit heuristic for the rectangular strip packing problem: An efficient implementation and the worst-case approximation ratio},
journal = {Computers \& Operations Research},
volume = {37},
number = {2},
pages = {325-333},
year = {2010},
issn = {0305-0548},
doi = {https://doi.org/10.1016/j.cor.2009.05.008},
url = {https://www.sciencedirect.com/science/article/pii/S030505480900149X},
author = {Shinji Imahori and Mutsunori Yagiura}
}

@InProceedings{heur-benders-1,
author="Raidl, G{\"u}nther R.
and Baumhauer, Thomas
and Hu, Bin",
editor="Blesa, Maria J.
and Blum, Christian
and Vo{\ss}, Stefan",
title="Speeding Up Logic-Based Benders' Decomposition by a Metaheuristic for a Bi-Level Capacitated Vehicle Routing Problem",
booktitle="Hybrid Metaheuristics",
year="2014",
publisher="Springer International Publishing",
address="Cham",
pages="183--197",
isbn="978-3-319-07644-7"
}

@article{heur-benders-2,
title = {Boosting an Exact Logic-Based Benders Decomposition Approach by Variable Neighborhood Search},
journal = {Electronic Notes in Discrete Mathematics},
volume = {47},
pages = {149-156},
year = {2015},
note = {The 3rd International Conference on Variable Neighborhood Search (VNS'14)},
issn = {1571-0653},
doi = {https://doi.org/10.1016/j.endm.2014.11.020},
url = {https://www.sciencedirect.com/science/article/pii/S1571065314000626},
author = {Günther R. Raidl and Thomas Baumhauer and Bin Hu}
}

@article{grus-placement,
title = {Automated placement of analog integrated circuits using priority-based constructive heuristic},
journal = {Computers \& Operations Research},
volume = {167},
pages = {106643},
year = {2024},
issn = {0305-0548},
doi = {https://doi.org/10.1016/j.cor.2024.106643},
url = {https://www.sciencedirect.com/science/article/pii/S0305054824001151},
author = {Josef Grus and Zdeněk Hanzálek}
}

@article{topo,
author = {Park, Young Woong and Klabjan, Diego},
title = {Bayesian network learning via topological order},
year = {2017},
issue_date = {January 2017},
publisher = {JMLR.org},
volume = {18},
number = {1},
issn = {1532-4435},
journal = {J. Mach. Learn. Res.},
month = jan,
pages = {3451–3482},
numpages = {32}
}

@inproceedings{hierarchical-planning-survey,
  title     = {A Survey on Hierarchical Planning – One Abstract Idea, Many Concrete Realizations},
  author    = {Bercher, Pascal and Alford, Ron and Höller, Daniel},
  booktitle = {Proceedings of the Twenty-Eighth International Joint Conference on
               Artificial Intelligence, {IJCAI-19}},
  publisher = {International Joint Conferences on Artificial Intelligence Organization},
  pages     = {6267--6275},
  year      = {2019},
  month     = {7},
  doi       = {10.24963/ijcai.2019/875},
  url       = {https://doi.org/10.24963/ijcai.2019/875},
}

@ARTICLE{fog-hierarchy,
  author={Peixoto, Maycon Leone Maciel and Genez, Thiago A. L. and Bittencourt, Luiz F.},
  journal={IEEE Transactions on Services Computing}, 
  title={Hierarchical Scheduling Mechanisms in Multi-Level Fog Computing}, 
  year={2022},
  volume={15},
  number={5},
  pages={2824-2837},
  doi={10.1109/TSC.2021.3079110}}

@InProceedings{edge-hierarchy,
author="Lane, Peter
and Helian, Na
and Bodla, Muhammad Haad
and Zheng, Minghua
and Moggridge, Paul",
editor="Jim{\'e}nez Laredo, Juan Luis
and Hidalgo, J. Ignacio
and Babaagba, Kehinde Oluwatoyin",
title="Dynamic Hierarchical Structure Optimisation for Cloud Computing Job Scheduling",
booktitle="Applications of Evolutionary Computation",
year="2022",
publisher="Springer International Publishing",
address="Cham",
pages="301--316"
}

@article{nested-column-cutting,
author = {Vanderbeck, Fran\c{c}ois},
title = {A Nested Decomposition Approach to a Three-Stage, Two-Dimensional Cutting-Stock Problem},
journal = {Management Science},
volume = {47},
number = {6},
pages = {864-879},
year = {2001},
doi = {10.1287/mnsc.47.6.864.9809},
URL = {https://doi.org/10.1287/mnsc.47.6.864.9809}
}

@article{nested-column-vrp,
title = {Nested branch-and-price-and-cut for vehicle routing problems with multiple resource interdependencies},
journal = {European Journal of Operational Research},
volume = {276},
number = {2},
pages = {549-565},
year = {2019},
issn = {0377-2217},
doi = {https://doi.org/10.1016/j.ejor.2019.01.041},
url = {https://www.sciencedirect.com/science/article/pii/S0377221719300761},
author = {Christian Tilk and Michael Drexl and Stefan Irnich}
}

@article{nested-bd-disaster,
author = {Penghui Guo and Zhijie Sasha Dong and Jianjun Zhu},
title = {Nested logic-based Benders decomposition for disaster preparedness planning with horizontal coordination},
journal = {IISE Transactions},
volume = {0},
number = {0},
pages = {1--33},
year = {2025},
publisher = {Taylor \& Francis},
doi = {10.1080/24725854.2025.2491495},
URL = {https://doi.org/10.1080/24725854.2025.2491495}
}

@article{mo-nested-bd,
title = {Electrification of transportation: A hybrid Benders/SDDP algorithm for optimal charging station trading},
journal = {International Journal of Hydrogen Energy},
volume = {89},
pages = {1060-1074},
year = {2024},
issn = {0360-3199},
doi = {https://doi.org/10.1016/j.ijhydene.2024.09.345},
url = {https://www.sciencedirect.com/science/article/pii/S0360319924040709},
author = {Farnaz Sohrabi and Mohammad Rohaninejad and Július Bemš and Zdeněk Hanzálek}
}

@article{nested-cyril,
title = {Nested logic-based Benders decomposition for an integrated home healthcare problem},
journal = {European Journal of Operational Research},
volume = {328},
number = {1},
pages = {32-48},
year = {2026},
issn = {0377-2217},
doi = {https://doi.org/10.1016/j.ejor.2025.06.006},
url = {https://www.sciencedirect.com/science/article/pii/S0377221725004758},
author = {Abdalrahman Algendi and Sebastián Urrutia and Lars Magnus Hvattum and Rafael A. Melo}
}

@misc{cpsatlp,
  title = {CP-SAT},
  version = { v9.12 },
  author = {Laurent Perron and Frédéric Didier},
  organization = {Google},
  url = {https://developers.google.com/optimization/cp/cp_solver/},
  year = { 2025 }
}

@article{NOVAK2019687,
title = {Scheduling with uncertain processing times in mixed-criticality systems},
journal = {European Journal of Operational Research},
volume = {279},
number = {3},
pages = {687-703},
year = {2019},
issn = {0377-2217},
doi = {https://doi.org/10.1016/j.ejor.2019.05.038},
url = {https://www.sciencedirect.com/science/article/pii/S0377221719304680},
author = {Antonin Novak and Premysl Sucha and Zdenek Hanzalek}
}

@article{bilevelsurvey,
title = {Metaheuristics for bilevel optimization: A comprehensive review},
journal = {Computers \& Operations Research},
volume = {161},
pages = {106410},
year = {2024},
issn = {0305-0548},
doi = {https://doi.org/10.1016/j.cor.2023.106410},
url = {https://www.sciencedirect.com/science/article/pii/S0305054823002745},
author = {José-Fernando Camacho-Vallejo and Carlos Corpus and Juan G. Villegas}
}

@dataset{zenodo-grus,
  author       = {Grus, Josef and
                  Hanzalek, Zdenek and
                  Artigues, Christian and
                  Briand, Cyrille and
                  Hebrard, Emmanuel},
  title        = {Instances for the two-dimensional hierarchical
                   packing (2DHRP) problem
                  },
  month        = dec,
  year         = 2025,
  publisher    = {Zenodo},
  doi          = {10.5281/zenodo.17876374},
  url          = {https://doi.org/10.5281/zenodo.17876374},
}

@misc{grus2026hrp,
  author       = {Grus, Josef},
  title        = {Hierarchical Rectangle Packing Solved by Multi-Level
                  Recursive Logic-based Benders Decomposition},
  year         = {2026},
  publisher    = {GitLab},
  url          = {https://gitlab.com/grusjose/hierarchical-rectangle-packing-solved-by-multi-level-recursive-logic-based-benders-decomposition}
}
